%% file: data__rarport_arxiv_v3.tex
\PassOptionsToPackage{x11names}{xcolor}
\documentclass[12pt,a4paper,twoside]{article}

\usepackage{array}
\usepackage{amsmath}
\usepackage{tabularx}
\usepackage{txfonts}
\usepackage{enumerate}
\usepackage{float}
\usepackage{xspace}
\usepackage[x11names]{xcolor}
\usepackage{multirow}

\usepackage{booktabs}

\usepackage{subcaption}

\usepackage{pdflscape}
\usepackage{graphicx}

\usepackage[T1]{fontenc}
\usepackage[cp1250]{inputenc}

\usepackage[margin=2cm]{geometry}

\usepackage[colorlinks=true,citecolor=red,linkcolor=Green4]{hyperref}

\usepackage{changes}

\usepackage{listings}
\colorlet{cpop}{Chartreuse3}
\colorlet{cmonika}{DarkOrchid1}
\colorlet{ckonrad}{cyan}

\definechangesauthor[name={Konrad}, color=ckonrad]{Konrad}
\definechangesauthor[name={Monika}, color=cmonika]{Monika}
\definechangesauthor[name={POP}, color=cpop]{POP}

\numberwithin{equation}{section}

\begin{document}
\pagestyle{myheadings}
\markright{Piotrowska,~M.J.\& Sakowski, K.: Analysis of the hospital network data}

\noindent\begin{tabular}{|p{\textwidth}}
	\Large\bf Analysis of the AOK Lower Saxony hospitalisation records data (years 2008 -- 2015)
 \\\vspace{0.01cm}
    \it Piotrowska, M.J.$^{\dagger,1}$ \& Sakowski, K$^{\dagger,*,2}$\\\vspace{0.02cm}
\it\small $^\dagger$Institute of Applied Mathematics and Mechanics,\\
\it\small University of Warsaw, Banacha 2, 02-097 Warsaw, Poland\\\vspace{0.01cm}
\it\small $^*$Institute of High Pressure Physics,\\
\it\small Polish Academy of Sciences, 01-142 Warsaw, Sokolowska 29/37, Poland\\\vspace{0.01cm}
\small  $^1$\texttt{monika@mimuw.edu.pl}, $^2$\texttt{konrad@mimuw.edu.pl}\\
    \multicolumn{1}{|r}{\large\color{orange} 
    } \\
	\\
	\hline
\end{tabular}
\thispagestyle{empty}

\tableofcontents
\addtocontents{toc}{\protect\setcounter{tocdepth}{3}}
\noindent\begin{tabular}{p{\textwidth}}
	\\
	\hline
\end{tabular}
\vspace{2em}\\

\begin{abstract}
Multidrug-resistant Enterobacteriaceae (MDR-E) have become a major public health threat in many
European countries. While traditional infection control strategies primarily target the containment of intra-hospital transmission, there is growing evidence highlighting the importance of inter-hospital patient traffic for the spread of MDR-E within healthcare systems. 

Our aim is to propose a network model, which reflect patient traffic in healthcare system and thus provide the framework to study transmission dynamics of MDR-E and the effectiveness of infection control strategies to contain their spread within and potentially across healthcare systems. However, to do that first we need to analyse real patients data and base on that propose network model reflecting the complexity of the real hospital network connections and dynamics of patient transfers between healthcare facilities. 
\end{abstract}

\textbf{Keywords:}  
healthcare data analysis, overlapping data, healthcare network, network epidemiology

\section{Introduction}

	The discovery of antibiotics had revolutionized the medicine and it had an immense impact on the standard of living of all people in the world. It allowed to effectively treat many infectious diseases caused by bacteria and to reduce their spread.
	Unfortunately, susceptibility to an antibiotic degrades with time, as the bacteria often mutate, and the evolution promotes resistant organisms. This effect is substantially augumented by mass-scale usage of the antibiotics, especially due to their frequent misuse. Recently, this problem has intensified~\cite{PerspectMedChem-2014--25} leading to emergence of antibiotic-resistant bacteria, which are not susceptible to some or most of the currently known antibiotics~\cite{ArchMedRes-2005-36-697}. Furthermore the development of new antibiotic is limited and recently has slowed down due to restrictive policies~\cite{CurrOpinMicrobiol-2003-6-427} and being not economically-sound for the pharmaceutical companies~\cite{ClinMicrobiolInfect-2006-12-25}.

	An important part of this problem, which is subject to this study, corresponds to hospital-acquired infections~\cite{NewEnglJMed-2010-362-1804}. It has been observed that multidrug-resistant Enterobacteriaceae infection rates grown considerably in Europe within recent years. While there are some preventive counter-measures~\cite{InfectControlHospEpidemiol-2011-32-1064,JAntimicrobChemother-2008-62-1422}, they are focused mostly on the facility level, instead of network of facilities or whole healthcare system. Clearly, hospitals exchange patients directly due to transfers, but also indirectly due to for example geographical proximity. These patients provide means of pathogen transfer between the healthcare facilities~\cite{Donker2012}. It is therefore beneficial, both in the sense of public health and economy, to focus on the multi-institutional or system-level preventive actions to control the pathogen spread~\cite{Ciccolini2013,Smith2005}, as the individual hospital goals may be contradictory~\cite{Real2005}.

	The simulations based on the network approach, where the healthcare facilities are considered as a base entities (network nodes), and the network edges corresponds to patient transfers, has already been successfully performed \cite{Donker2012,AmerJPublHealth-2011-101-707,Donker2017} showing that the cooperative approach may lead to better results than individual approach, at least in case of methicllinin-resistant \emph{Staphylococcus aureus}~\cite{Lee2012}.

	The first step to perform such simulations is to build a realistic healthcare facility network. Unfortunately even in well-developed countries with extensive health care the (anonymized) patient transfer databases are not freely available. However, it may be possible, to use a database of patient admission/discharge to reconstruct the network connection between the healthcare facilities.
	
	In this paper, we discuss the usage of such anonymized database, provided by AOK Lower Saxony, in context of derivation of the computer model of a regional healthcare system.
	The paper is organized as follows. In Sections~\ref{sec:description} and~\ref{sec:tools} we briefly describe the database under consideration and the utilities used for its analysis. Then the analysis is performed as presented in Section~\ref{sec:analysis}. Finally, in Section~\ref{sec:network} we recapitulate the main problems and we comment on building the network model from the data provided.

\section{Description of data set}
\label{sec:description}
 
In the presented study, we consider anonymized patients data set provided by AOK Lower Saxony -- a healthcare provider in a large federal state in Germany. Data set consist of 5\,254\,492 hospitalisation records of 1\,673\,247  patients (registered in AOK Lower Saxony) during the years 2008-2015. Data base consists in particular of the following information: patient anonymized ID, gender, anonymized healthcare facility ID, region code of healthcare facility, day of the admission, day of discharge, diagnosis (international ICD 10 code and numeric code).
It should be mentioned that, due to the personal data protection, the information on geographical location of healthcare facilities has been removed from the provided database.

%
%
%

Within provided data set we have found 4\,573\,584 hospital/healthcare facility stay records for the facilities located in Lower Saxony with the numeric diagnosis code and  680\,908 for other German federal states. Within these records, data related to healthcare facilities of any size were taken into account.
%
Records without the diagnosis code were omitted in our analysis. Among them, there are 257\,668 hospitalisation records for Lower Saxony facilities and 79\,102 records for the remaining parts of Germany. 

\section{Data analysis tools}
\label{sec:tools}

Provided data set consists of 5\,254\,492 records which makes it large. Thus, to perform the analysis described below we needed to process the data set provided by AOK Lower Saxony. From the admission and discharge data it was necessary to determine the duration of stays for each hospitalisation record. It was also necessary to group the records with the same patient identification number to detect existing overlaps (definition of the overlap will be provided in Subsection~\ref{sec:ovelaps}) in data set.

To do so we have developed code using Python programming language. 
In our code we used many widely available Python libraries, in particular \emph{matplotlib} library for visualization. Since the code was supposed to analyse the provided data, it was developed with possibly flexible approach, such that it is possible to extend the analysis depending on need. Optimization of code was the secondary goal, as this is not supposed to be a high-performance simulation code at this stage, but a statistic/analytical tool.

The basic idea standing behind our application is as follows. The AOK hospitalisation database, provided as a CSV (comma-separated value) file are parsed and transformed into Python classes, providing high-level subroutines. Then they are stored in a data structured (lists, dictionaries), which were used to determine overlapping records and their statistics. Application of dictionaries allowed us to perform efficient operations of searching among millions of records.

At this moment, the computation time for this application is about 1 hour for current database size and it uses up to 10 GB of RAM memory, as all the structures are stored in RAM during processing, and only the results are stored permanently. In future versions, however, we tend to use on-disk databases, like for example \emph{SQLite}, to overcome the problem of high memory consumption and to provide more flexibility in operating on the results.

\section{Data analysis}
\label{sec:analysis}
Within all the records we have found data for 757\,141 men with 1 up to 322 hospitalisations and 916\,106 women with 1 up to 195 within years 2008--2015.
The median and average number of entries per person for each sex are 2 and $\approx 3.3$ days for men and 2 and $\approx 3.4$ days for  women.

%

\subsection{Admissions}
\begin{figure}
	\centering
	\begin{subfigure}[t]{0.5\textwidth}
		\centering
		\includegraphics[height=6.5cm]{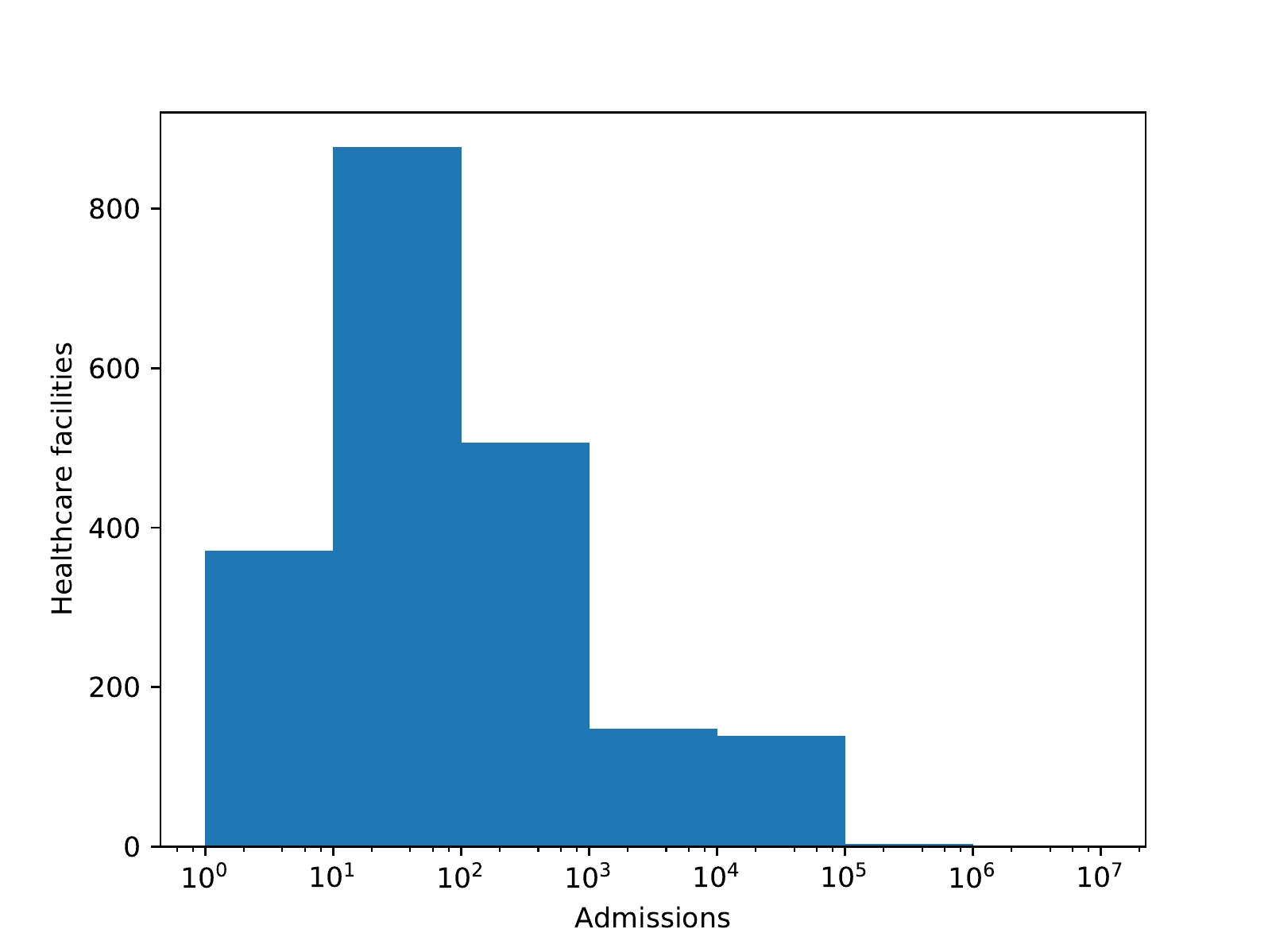}
		\caption{}
	\end{subfigure}%
	~ 
	\begin{subfigure}[t]{0.5\textwidth}
		\centering
		\includegraphics[height=6.5cm]{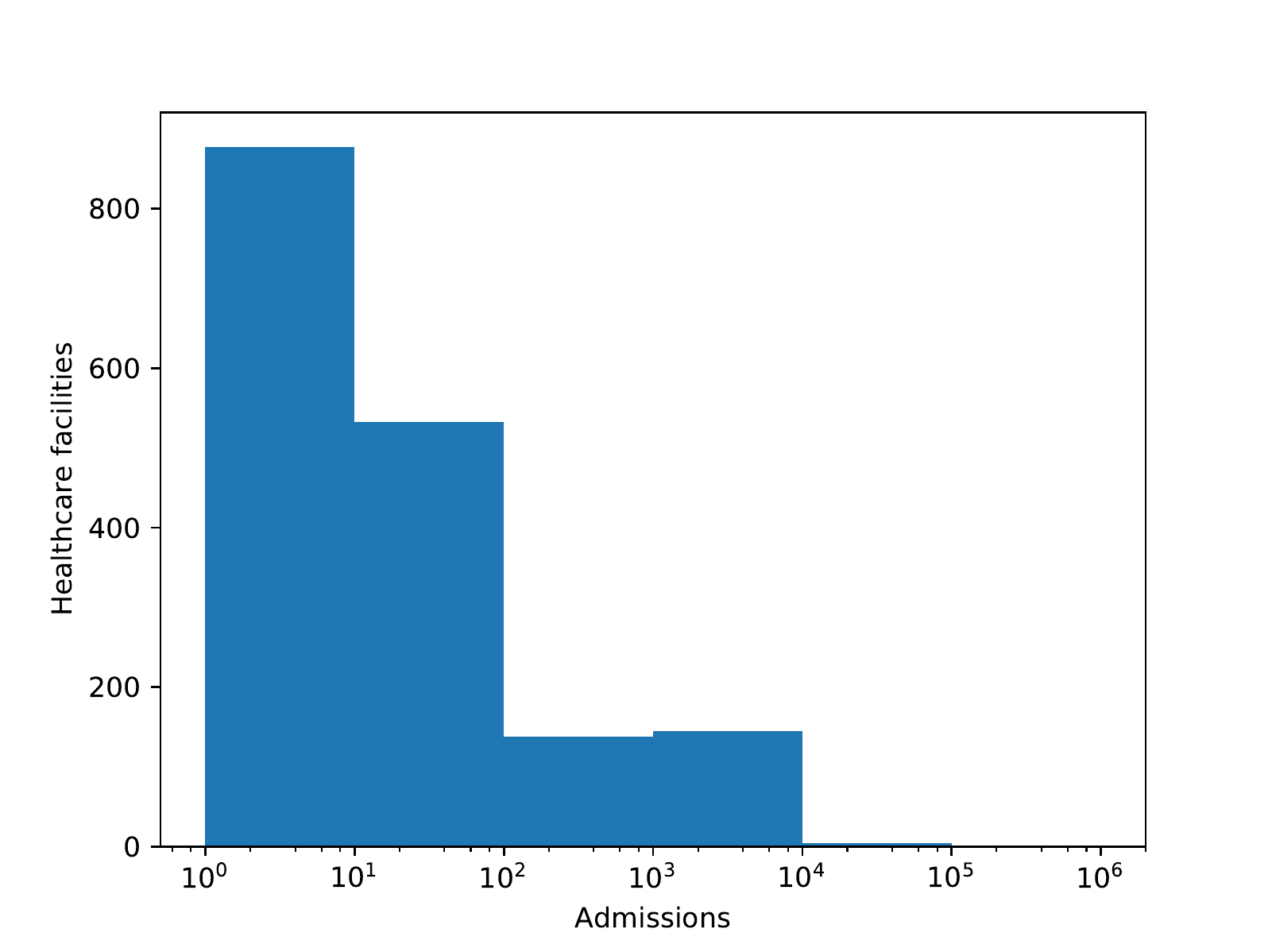}
		\caption{}
	\end{subfigure}  
	~
	\begin{subfigure}[t]{0.5\textwidth}
		\centering
		\includegraphics[height=6.5cm]{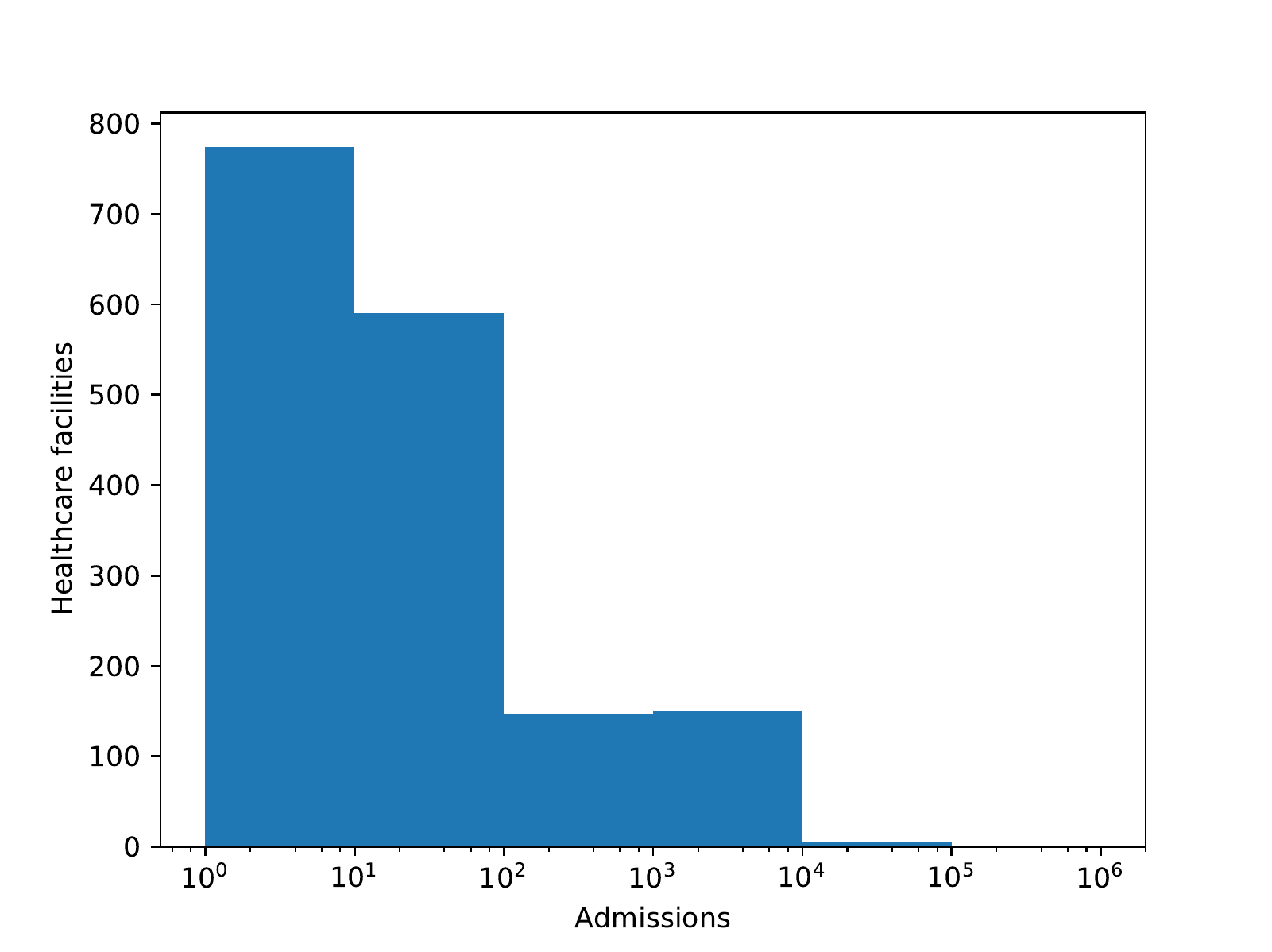}
		\caption{}
	\end{subfigure}%
	~ 
	\begin{subfigure}[t]{0.5\textwidth}
		\centering
		\includegraphics[height=6.5cm]{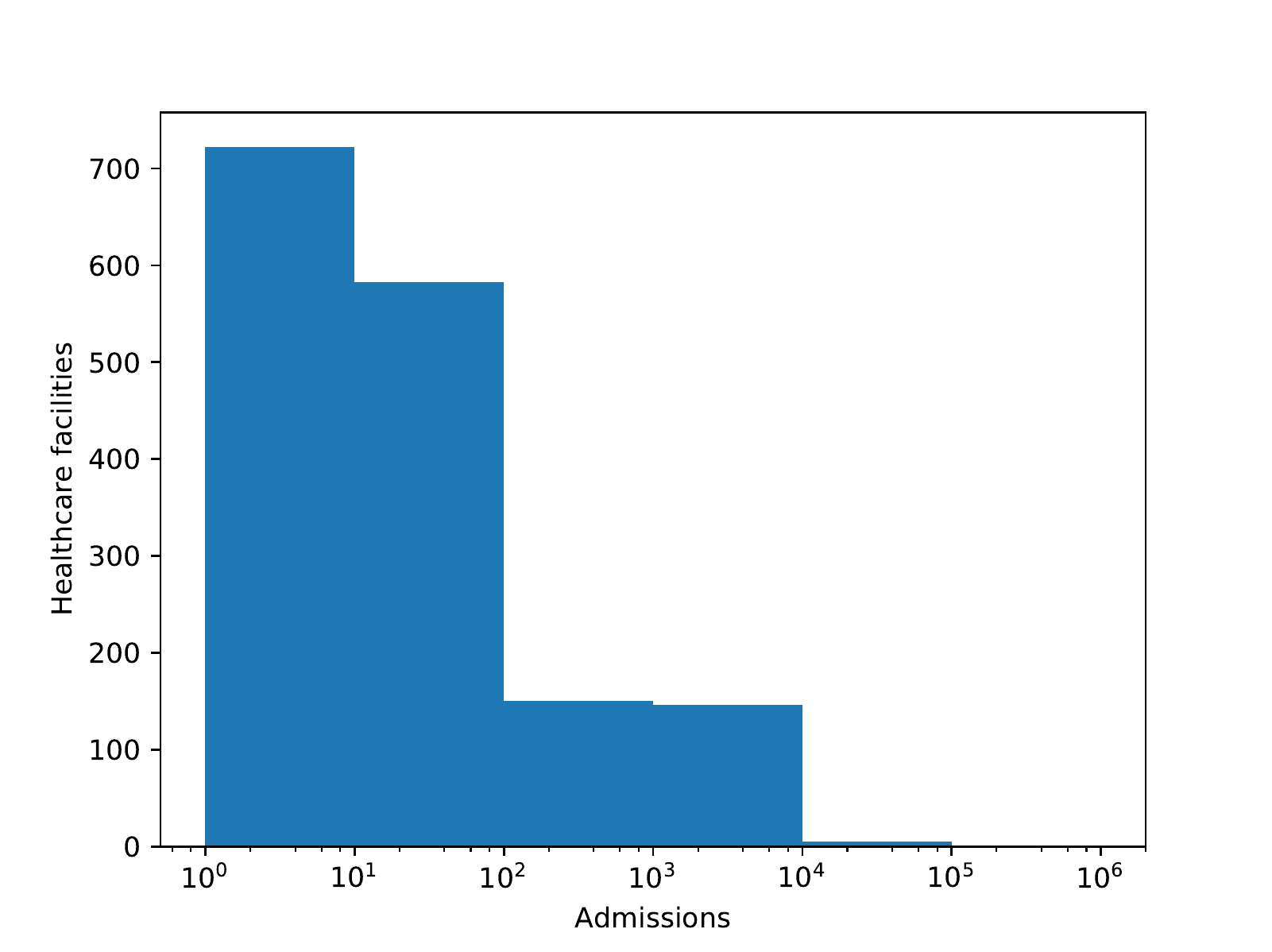}
		\caption{}
	\end{subfigure} 
	~
\begin{subfigure}[t]{0.5\textwidth}
	\centering
	\includegraphics[height=6.5cm]{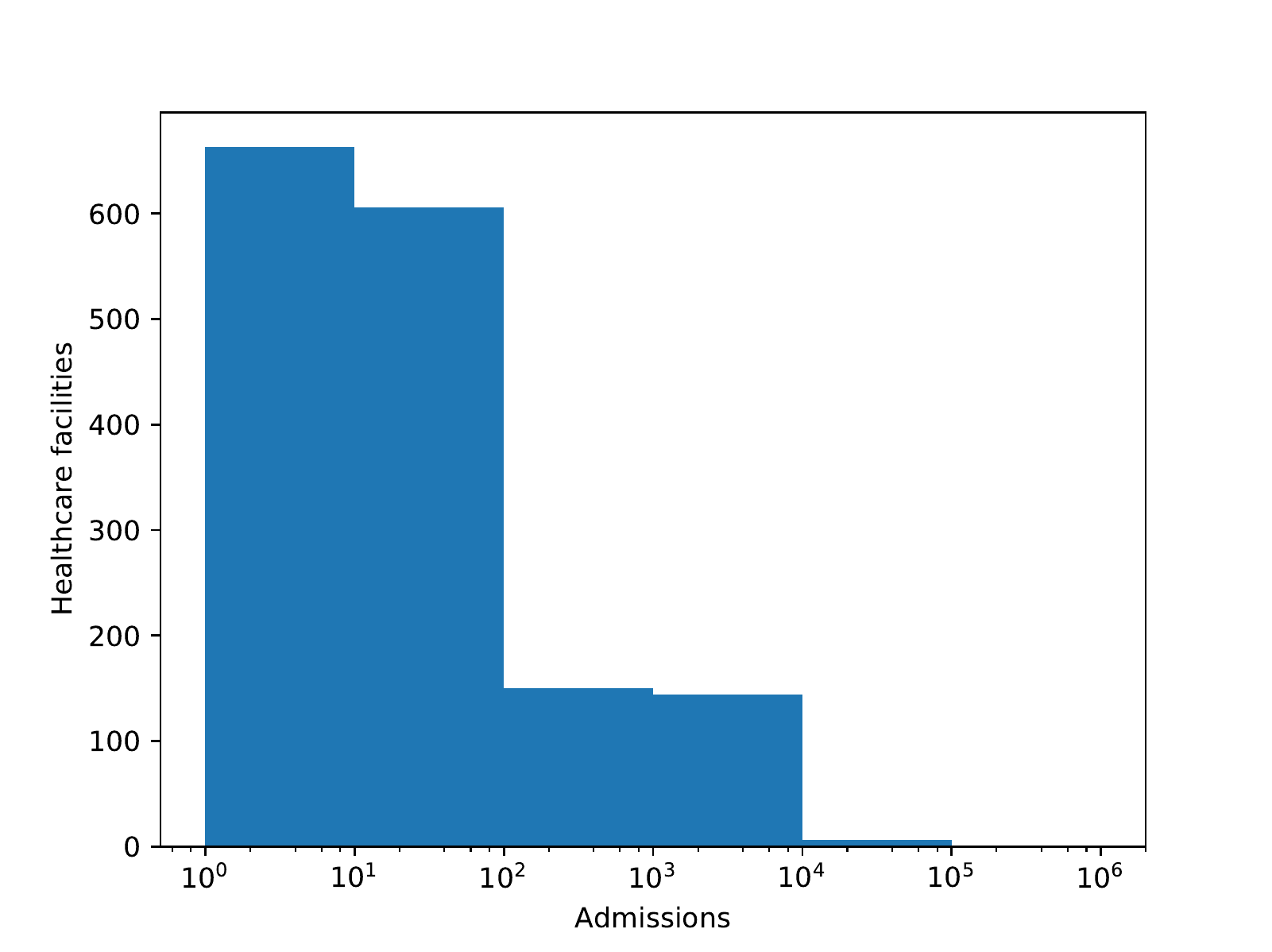}
	\caption{}
\end{subfigure}%
~ 
\begin{subfigure}[t]{0.5\textwidth}
	\centering
	\includegraphics[height=6.5cm]{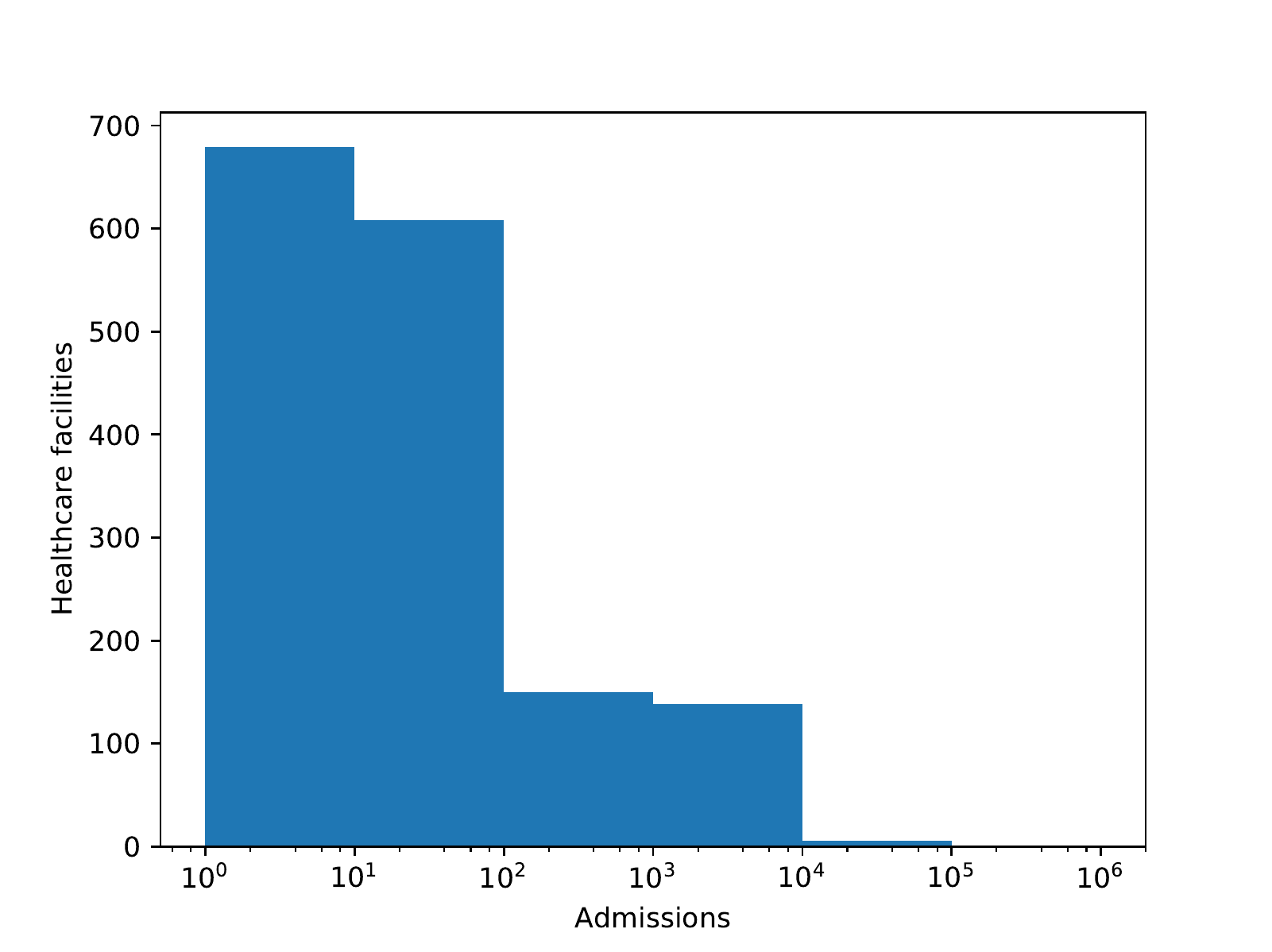}
	\caption{}
\end{subfigure} 
	\caption{Number of healthcare facilities having given number of admissions for all cases reported: (a)~within years 2008-2015, (b)~in 2008, (c)~in 2010, (d)~in 2012, (e)~in 2014 and (f)~in 2015.\label{fig:hosp:entries:all}}
\end{figure}

\begin{figure}
	\centering
	\begin{subfigure}[t]{0.5\textwidth}
		\centering
		\includegraphics[height=6.5cm]{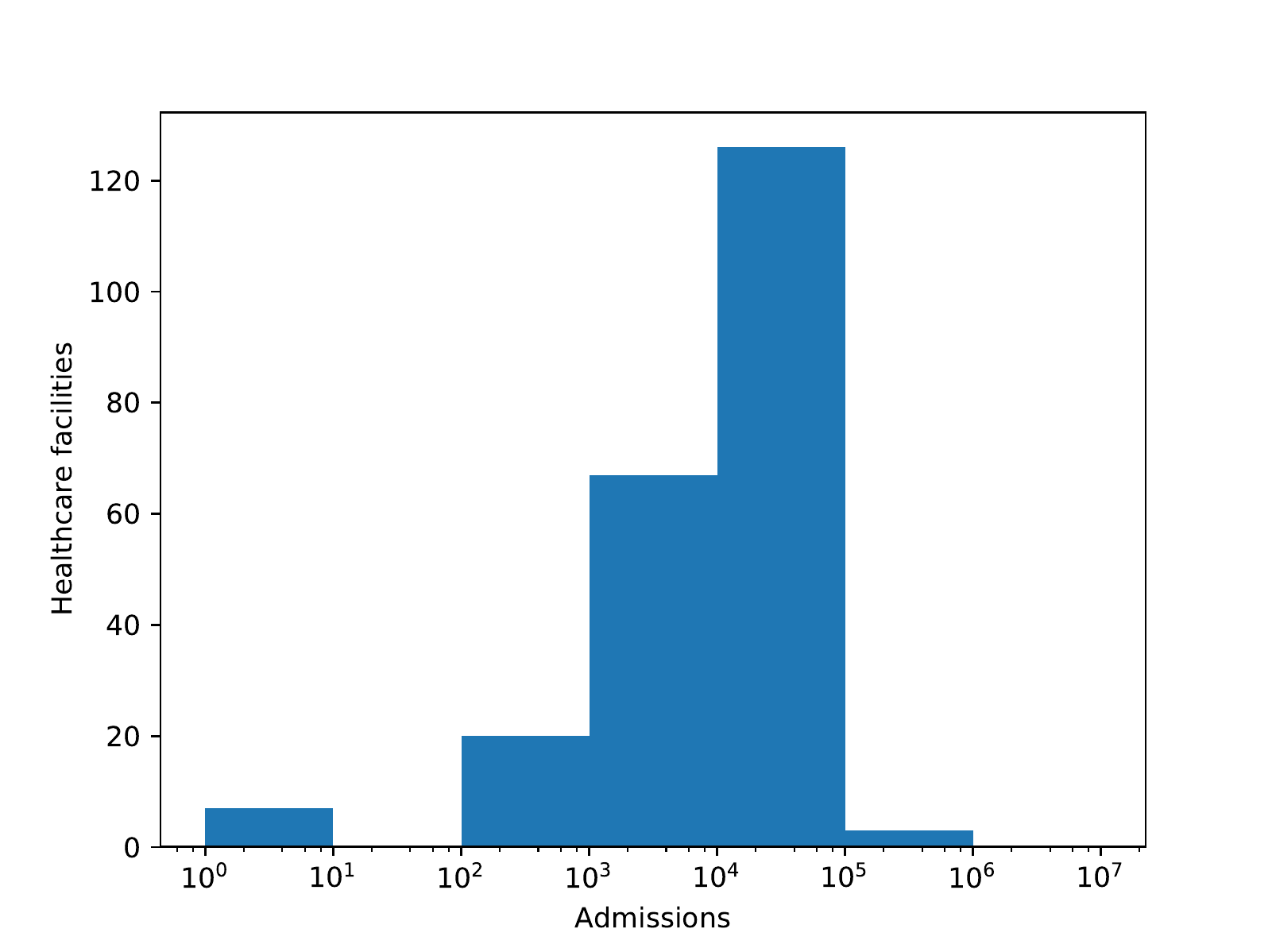}
		\caption{}
	\end{subfigure}%
	~ 
	\begin{subfigure}[t]{0.5\textwidth}
		\centering
		\includegraphics[height=6.5cm]{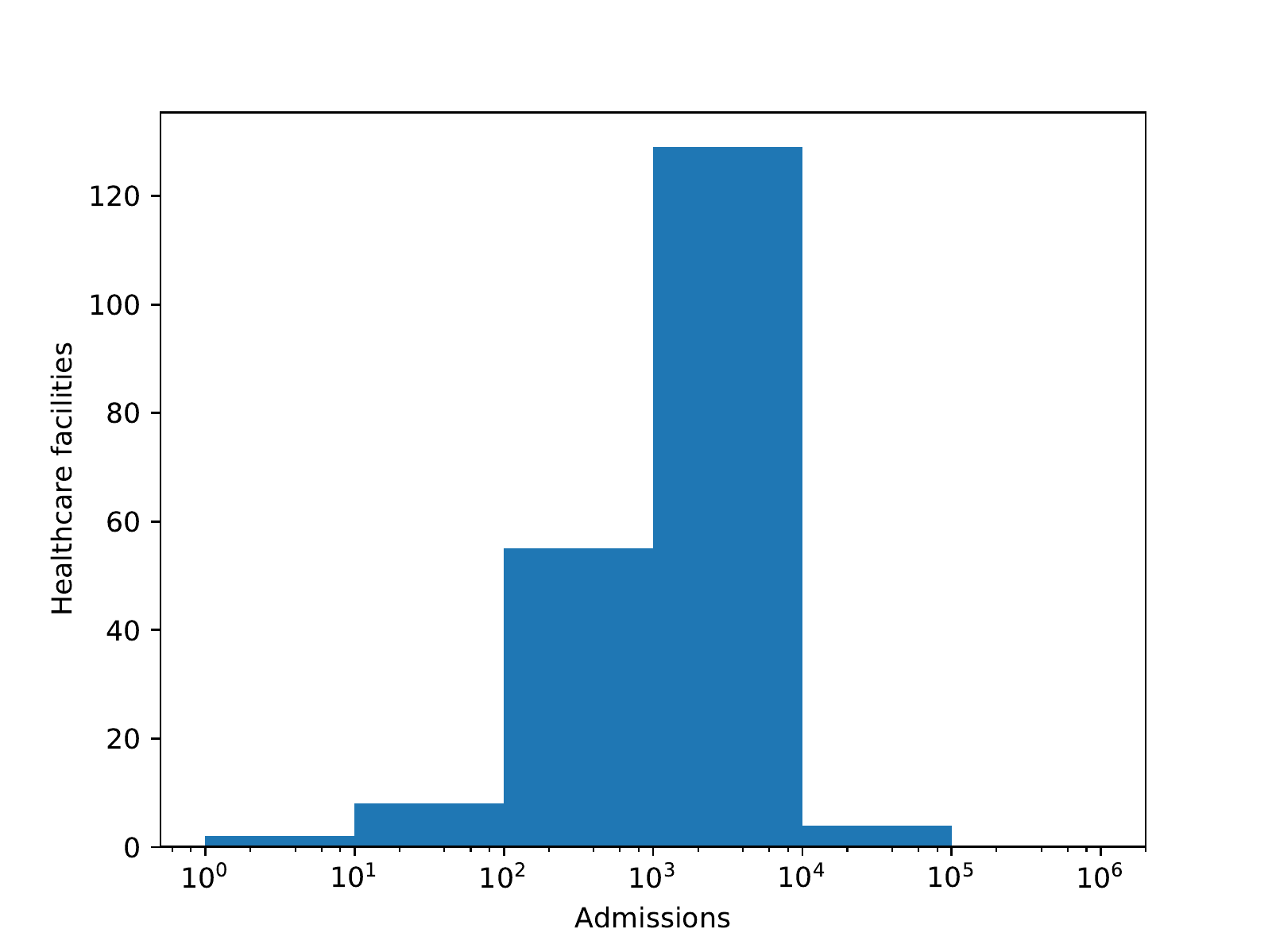}
		\caption{}
	\end{subfigure}  
	~
	\begin{subfigure}[t]{0.5\textwidth}
		\centering
		\includegraphics[height=6.5cm]{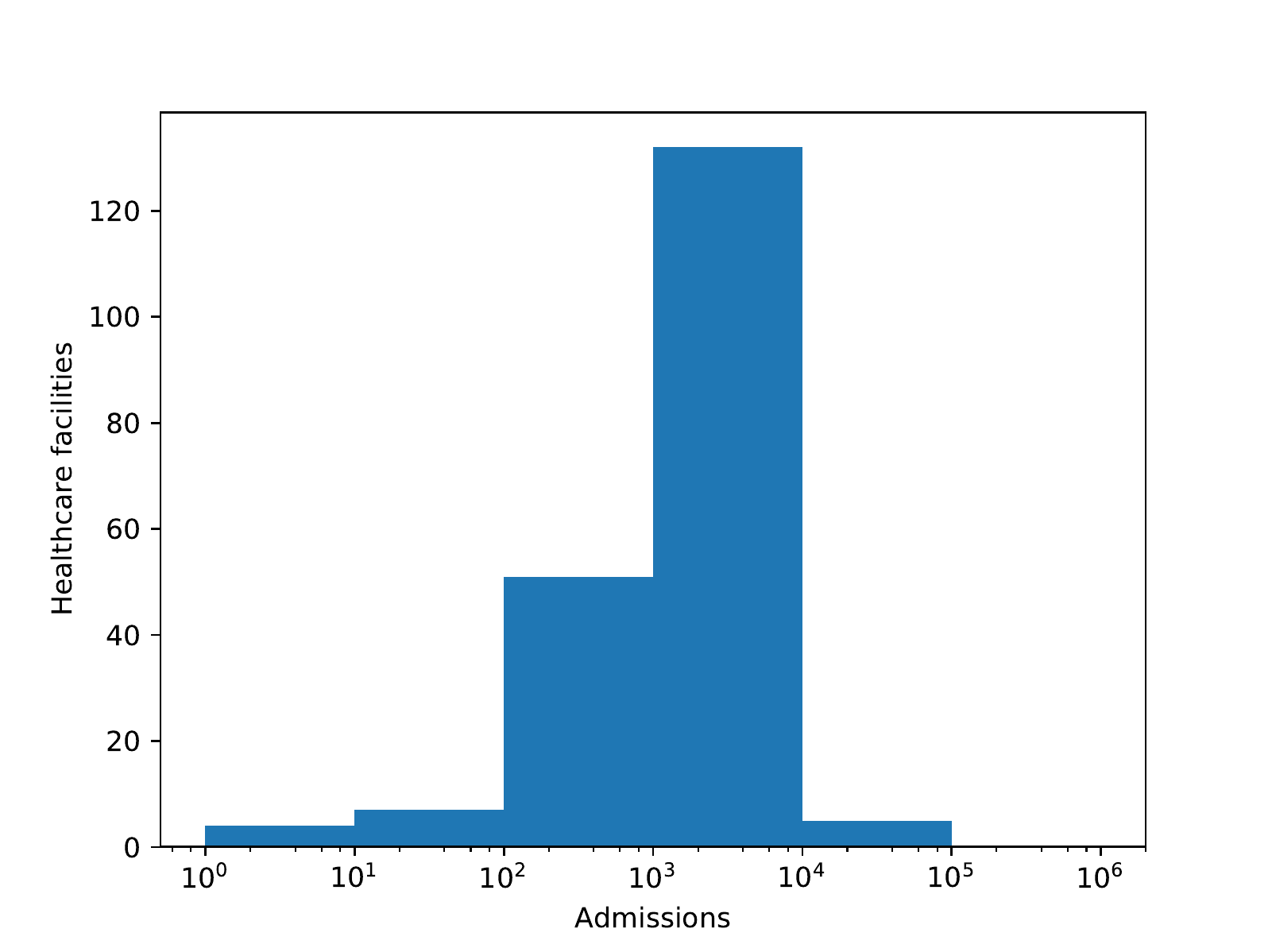}
		\caption{}
	\end{subfigure}%
	~ 
	\begin{subfigure}[t]{0.5\textwidth}
		\centering
		\includegraphics[height=6.5cm]{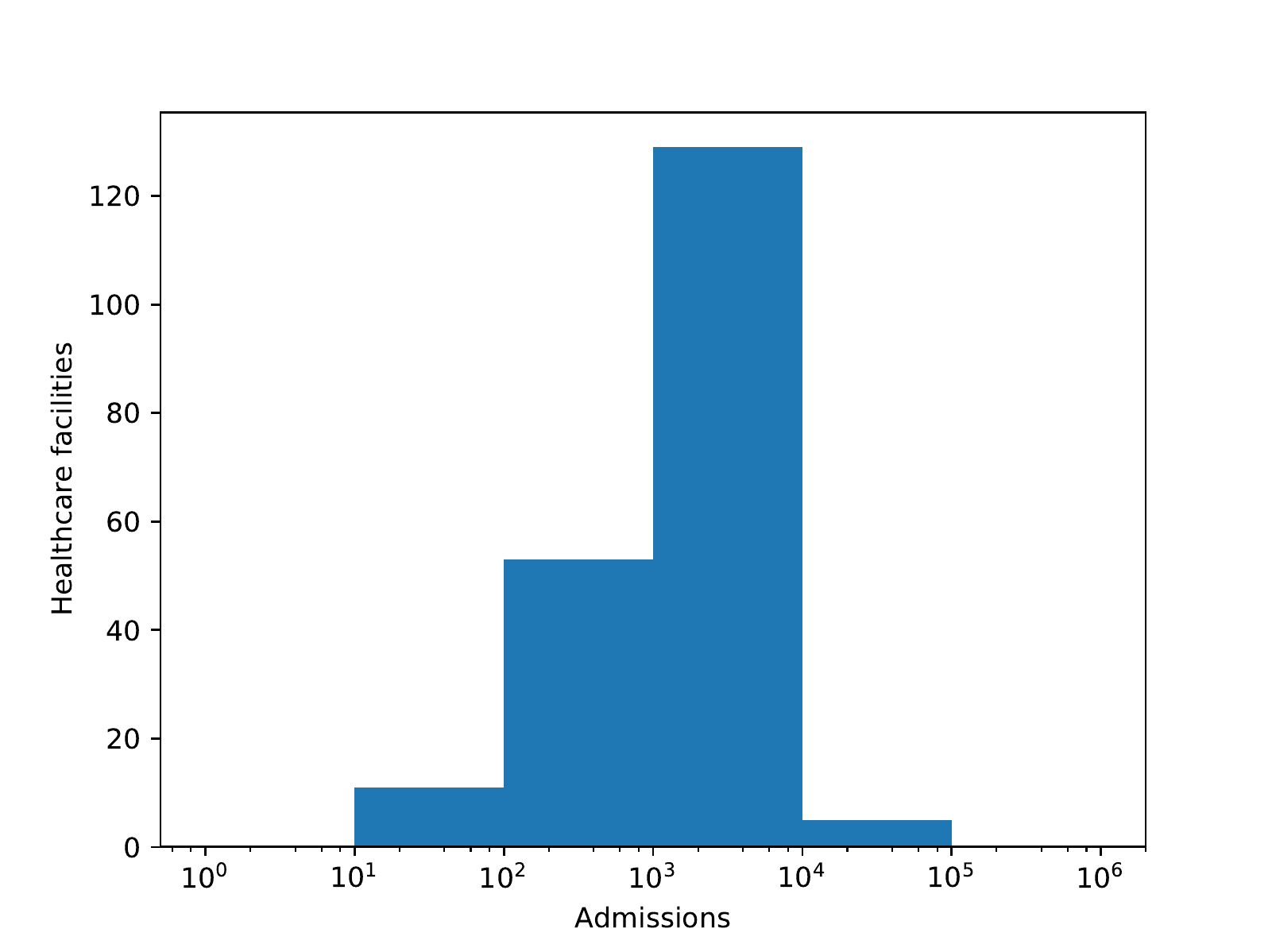}
		\caption{}
	\end{subfigure} 
	~
	\begin{subfigure}[t]{0.5\textwidth}
	\centering
	\includegraphics[height=6.5cm]{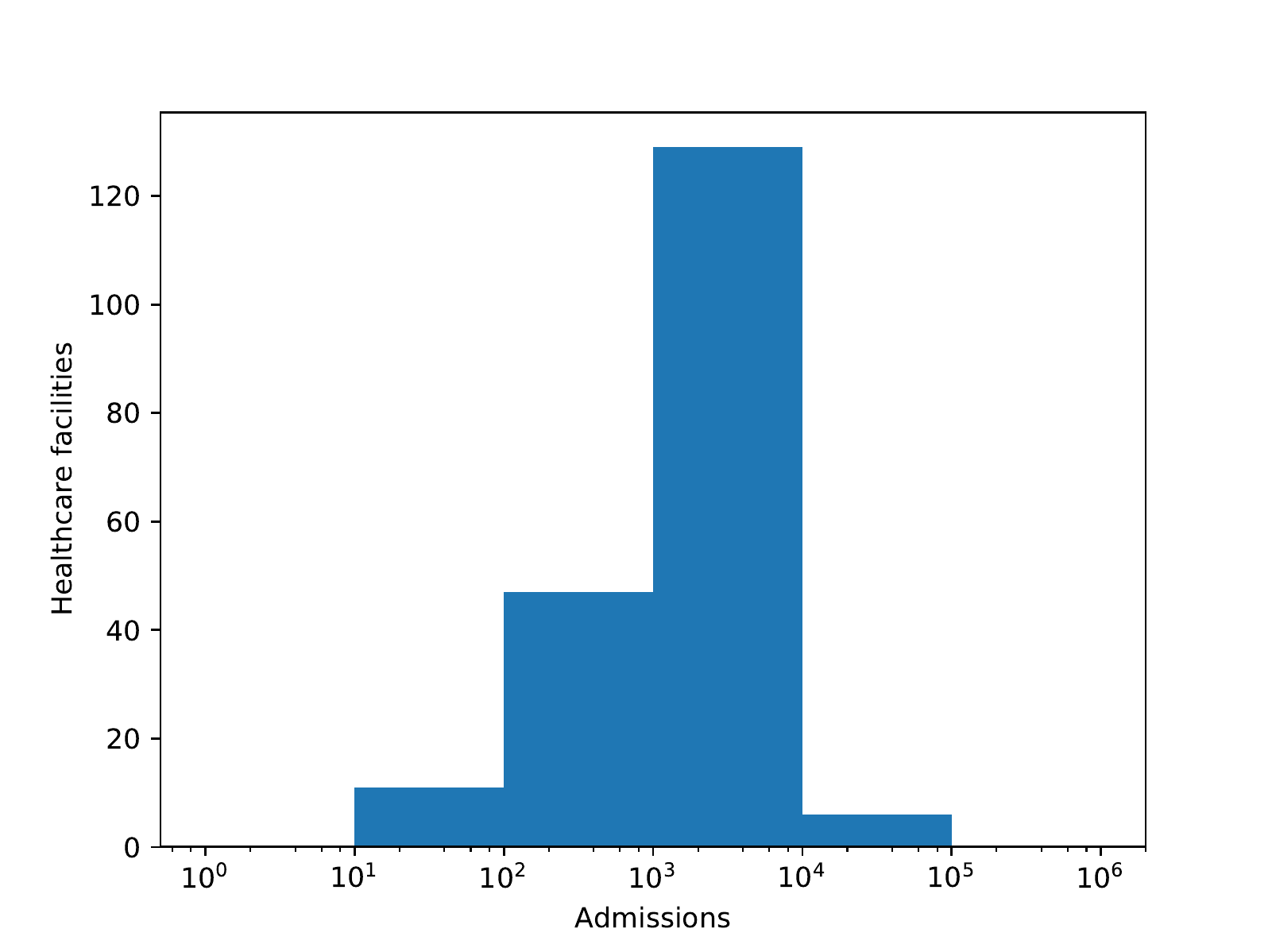}
	\caption{}
	\end{subfigure}%
	~ 
	\begin{subfigure}[t]{0.5\textwidth}
	\centering
	\includegraphics[height=6.5cm]{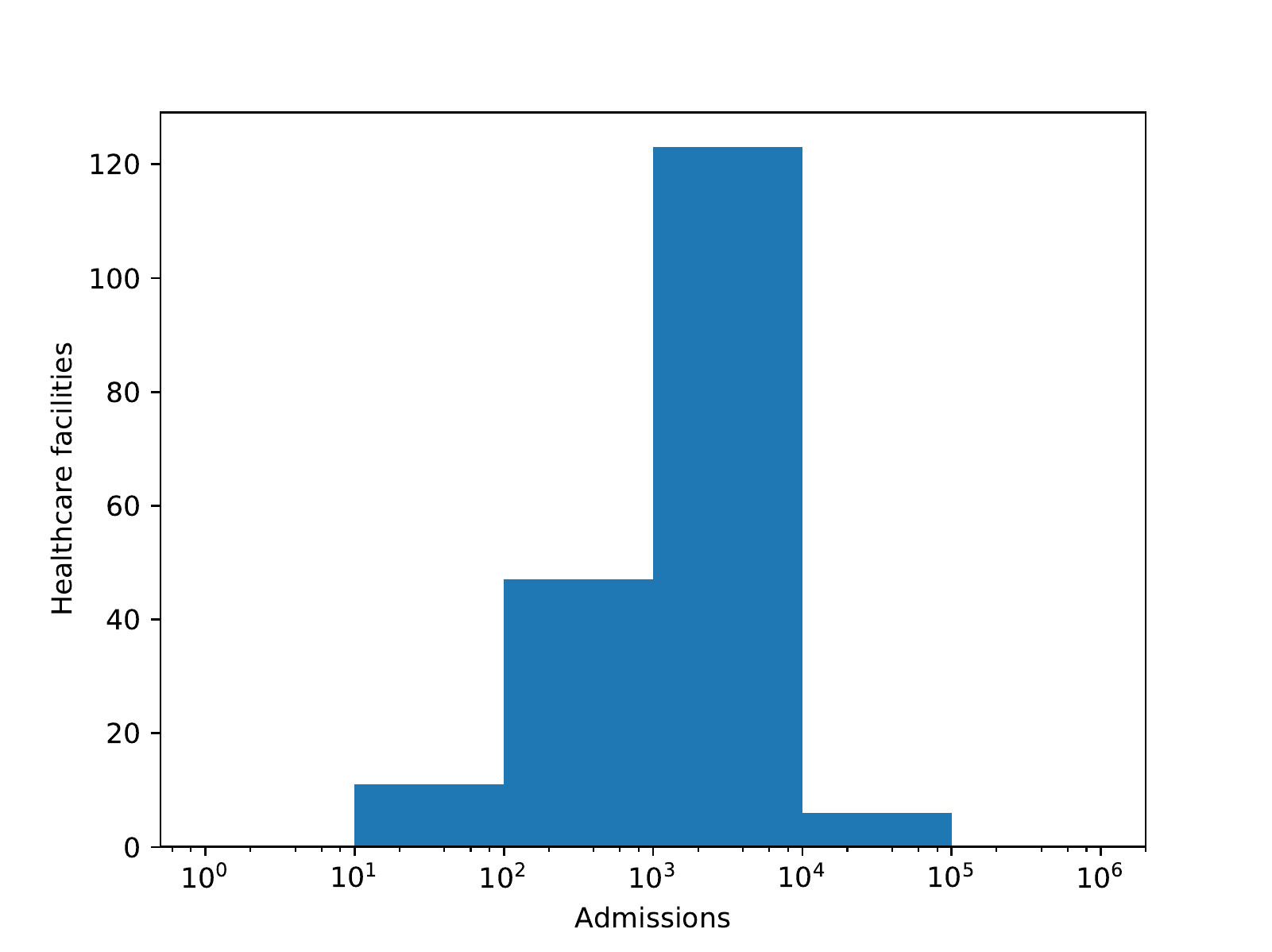}
	\caption{}
	\end{subfigure} 
	\caption{Number of healthcare facilities in Lower Saxony having given number of admissions reported: (a) within years 2008-2015, (b)~in 2008, (c)~in 2010, (d)~in 2012, (e)~in 2014 and (f)~in 2015.\label{fig:hosp:entries:hkbula03}}
	\end{figure}

Let focus on the characterisation of the healthcare facilities reported in the database.
From Figure~\ref{fig:hosp:entries:all} it is clear that the analysis of the whole data set indicates that the majority healthcare facilities within particular years had between 1 and 9 admissions within the considered time periods. For the collectively calculated years 2008--2015 the majority of healthcare facilities had between 10 and 99 reported admissions. However, if we restrict our analysis only to the healthcare facilities located in the Lower Saxony we observe different trend -- the majority healthcare facilities has between  10\,000 and 99\,999 admissions, for details see~Figure~\ref{fig:hosp:entries:hkbula03}. For separately treated years most Lower Saxony healthcare facilities had between 1\,000 and 9\,999 reported admissions. 

\subsection{Numbers of patients}

\begin{figure}
	\centering
	\begin{subfigure}[t]{0.5\textwidth}
		\centering
		\includegraphics[height=6.5cm]{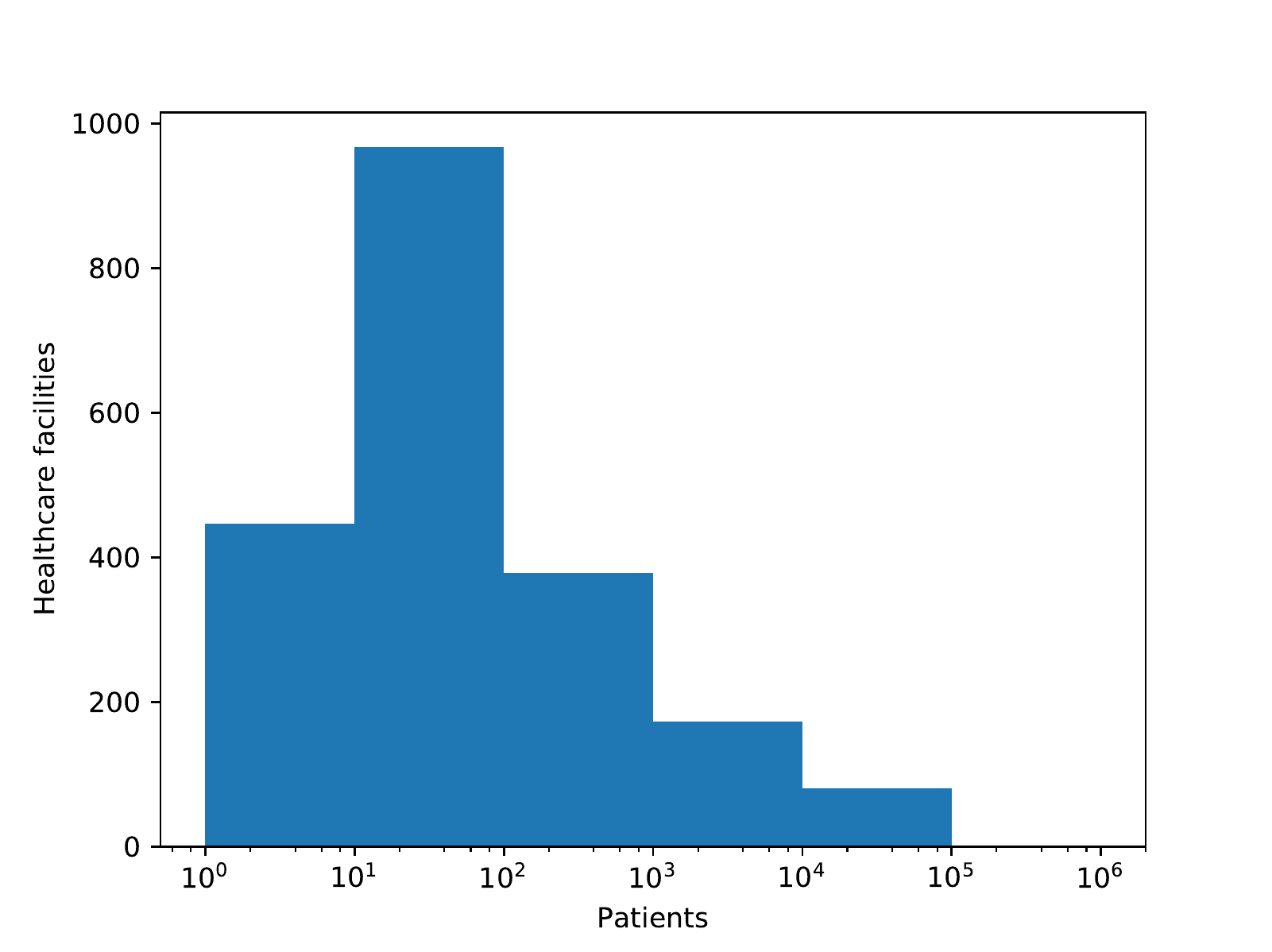}
		\caption{}
	\end{subfigure}%
	~ 
	\begin{subfigure}[t]{0.5\textwidth}
		\centering
		\includegraphics[height=6.5cm]{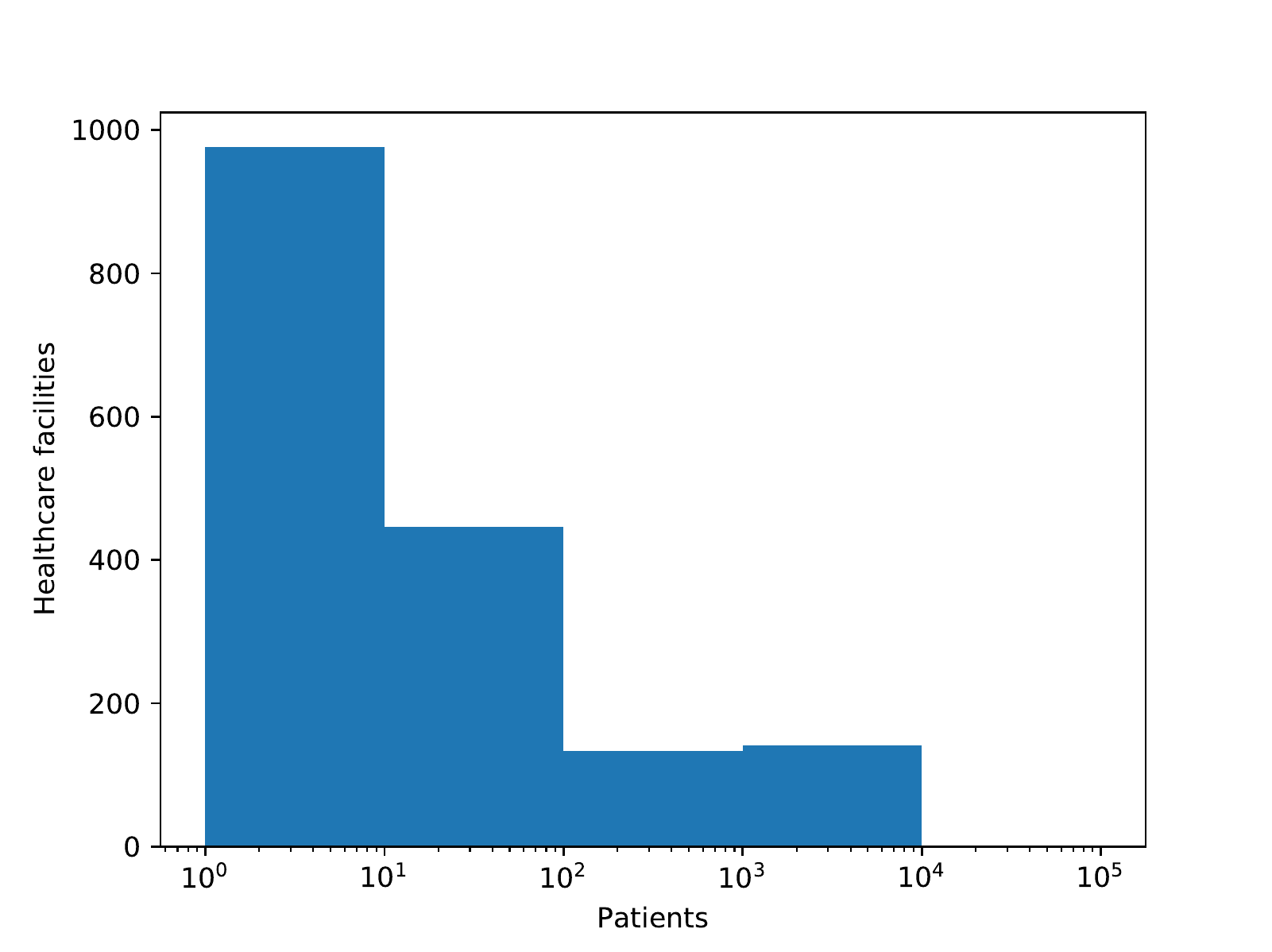}
		\caption{}
	\end{subfigure}  
	~
	\begin{subfigure}[t]{0.5\textwidth}
		\centering
		\includegraphics[height=6.5cm]{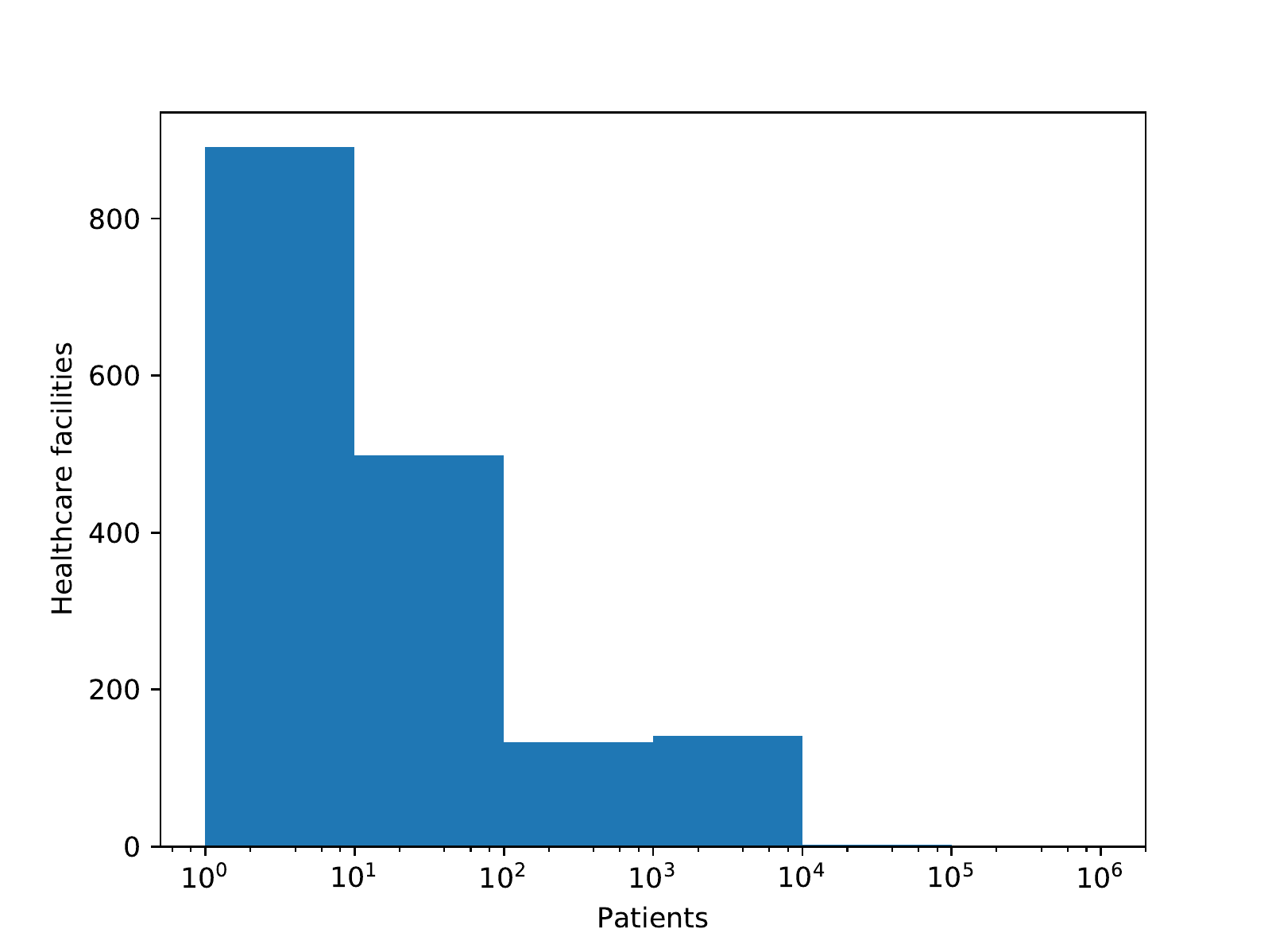}
		\caption{}
	\end{subfigure}%
	~ 
	\begin{subfigure}[t]{0.5\textwidth}
		\centering
		\includegraphics[height=6.5cm]{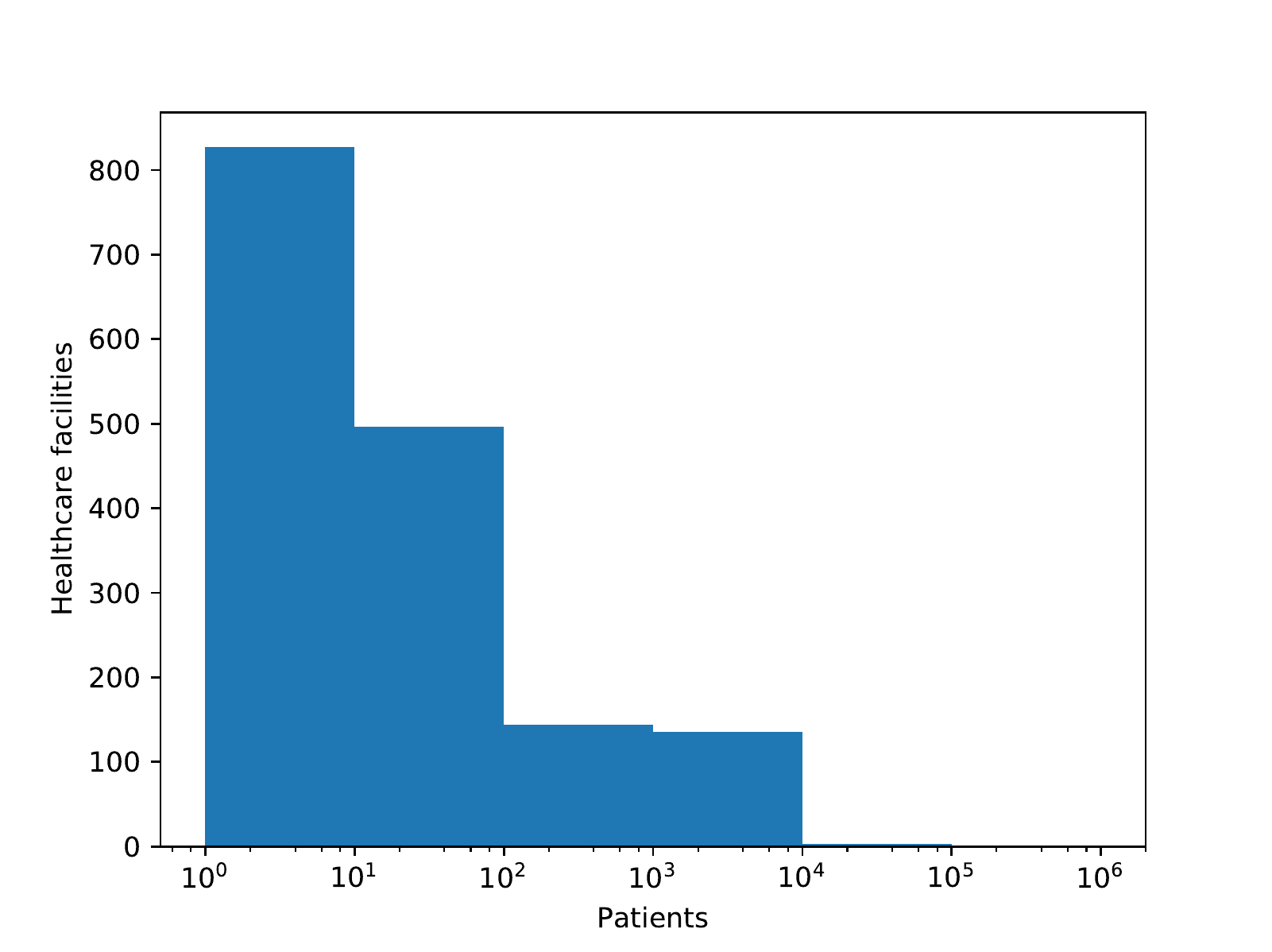}
		\caption{}
	\end{subfigure} 
	~
	\begin{subfigure}[t]{0.5\textwidth}
	\centering
	\includegraphics[height=6.5cm]{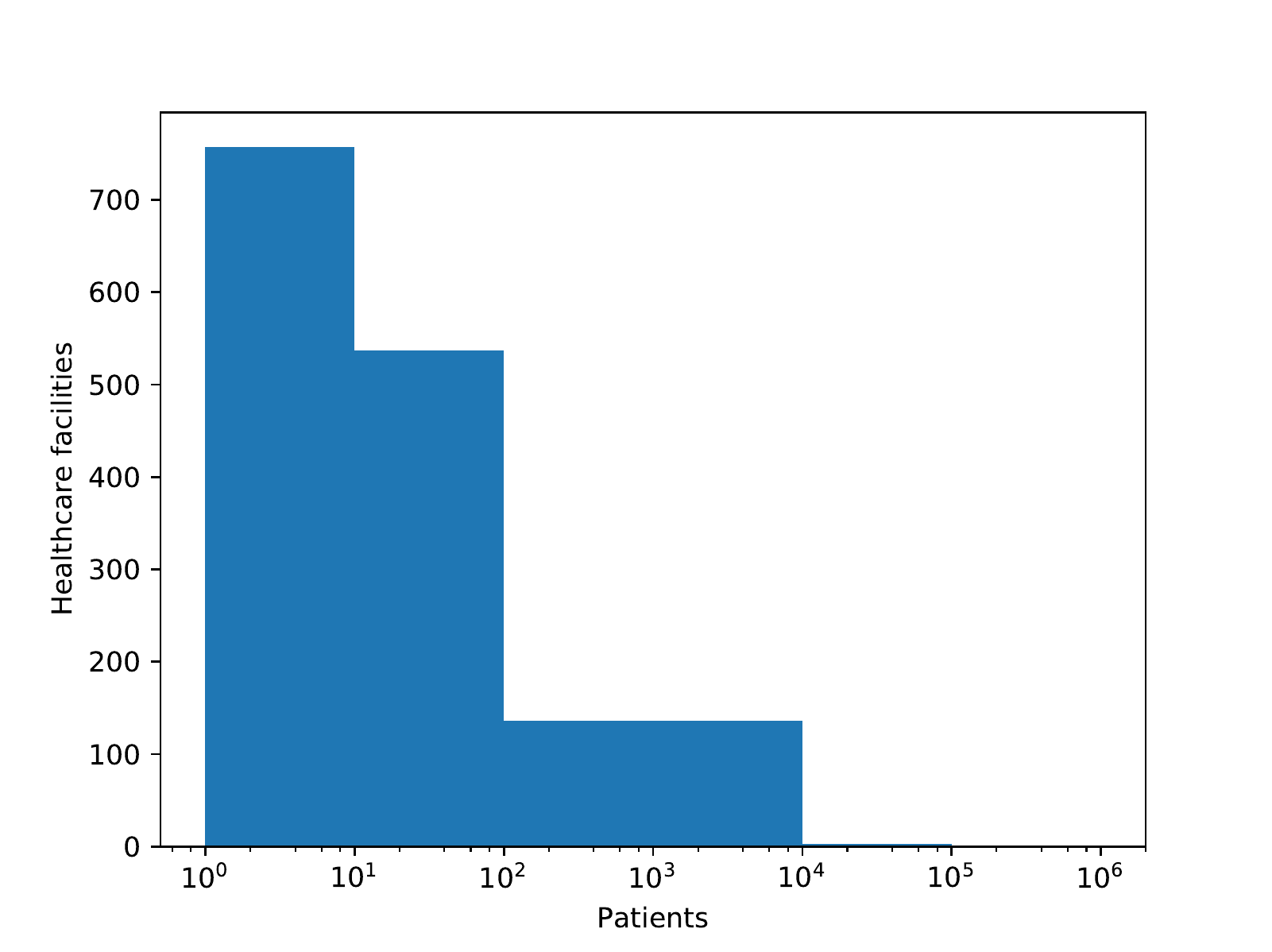}
	\caption{}
	\end{subfigure}%
	~ 
	\begin{subfigure}[t]{0.5\textwidth}
	\centering
	\includegraphics[height=6.5cm]{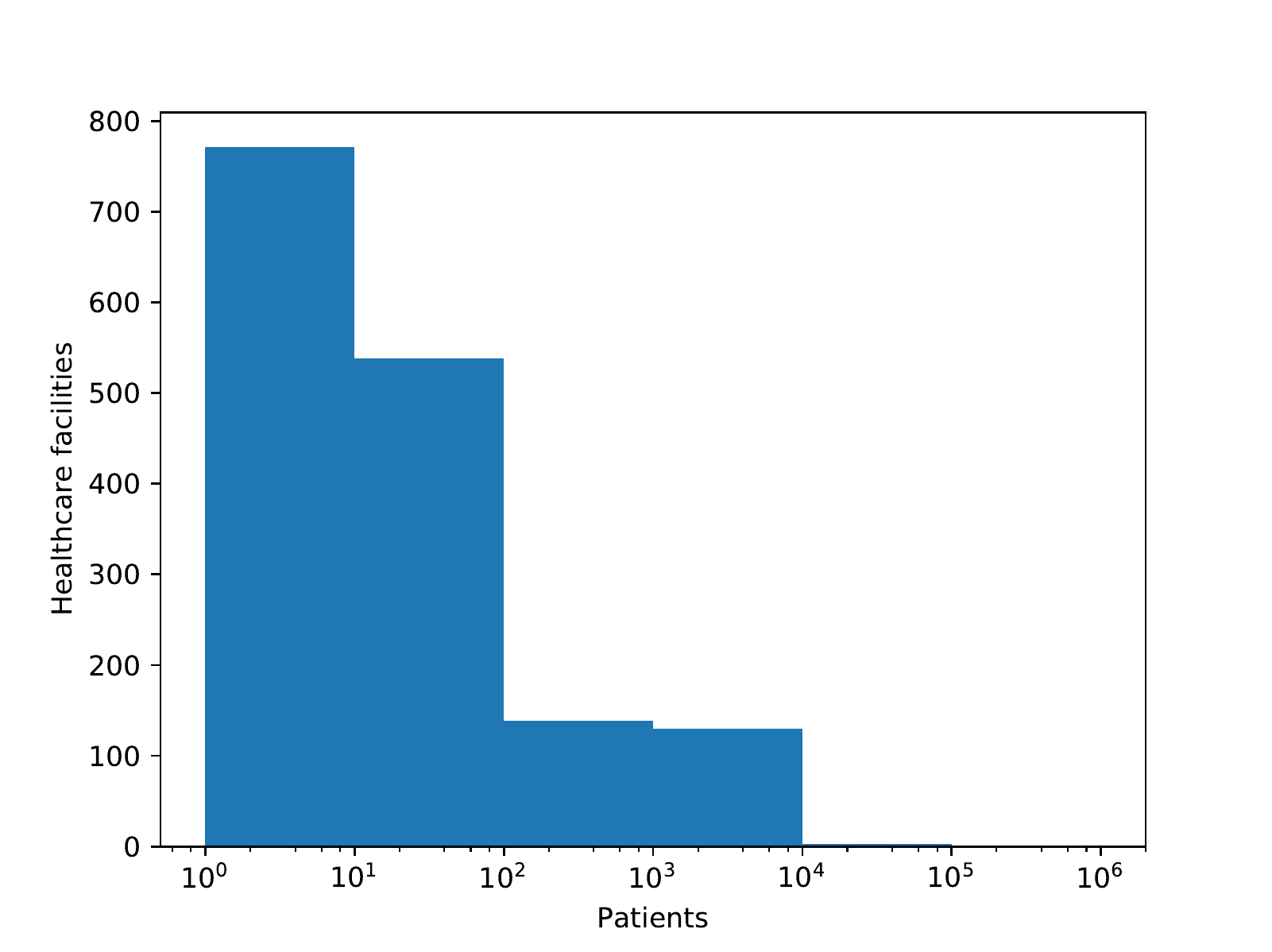}
	\caption{}
	\end{subfigure} 
	\caption{Numbers of healthcare facilities having given number of patients for all cases reported: (a)~within years 2008-2015, (b)~in 2008, (c)~in 2010, (d)~in 2012, (e)~in 2014 and (f)~in 2015.\label{fig:hosp:patients:all}}
\end{figure}

\begin{figure}
	\centering
	\begin{subfigure}[t]{0.5\textwidth}
		\centering
		\includegraphics[height=6.5cm]{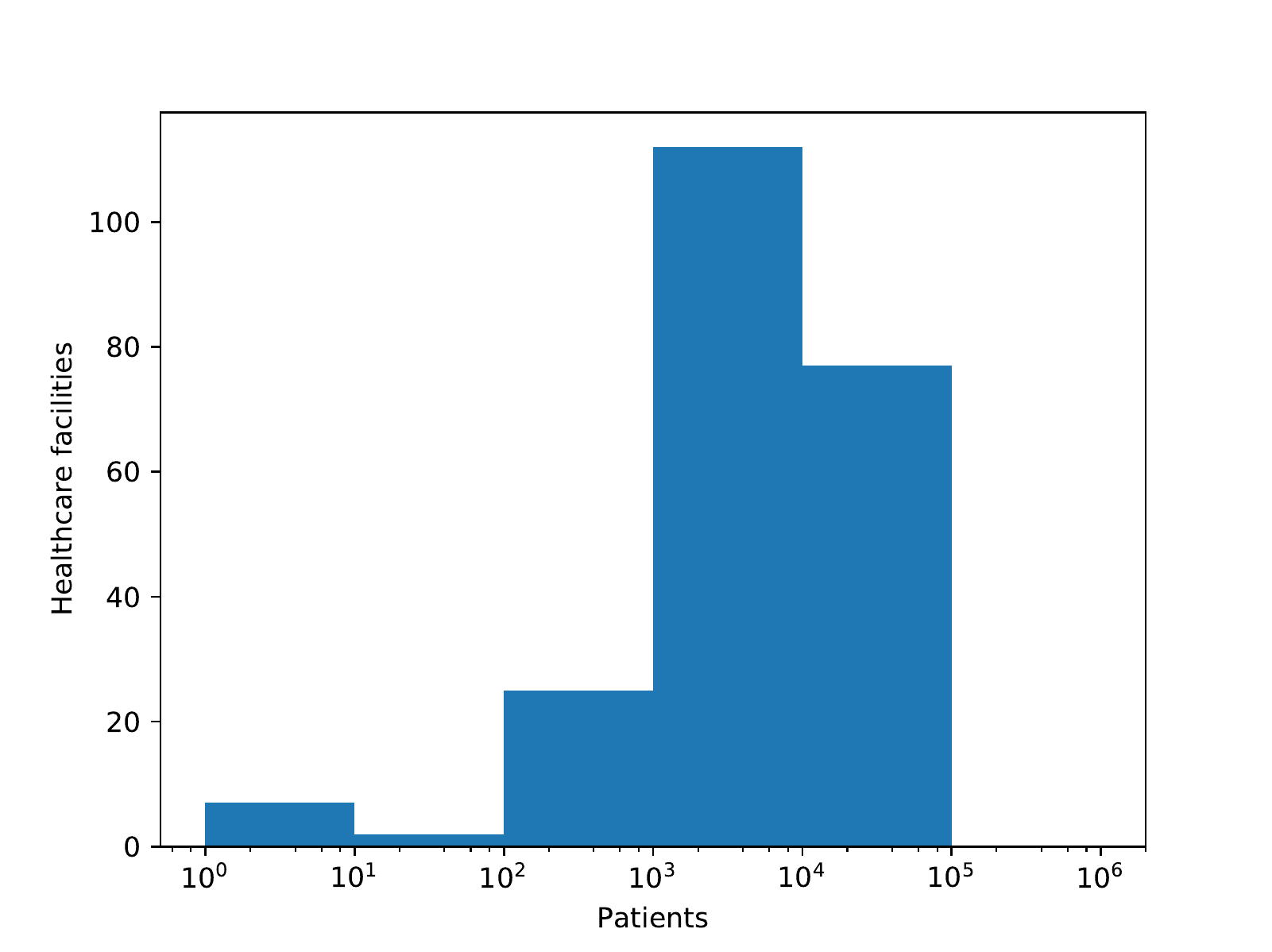}
		\caption{}
	\end{subfigure}%
	~ 
	\begin{subfigure}[t]{0.5\textwidth}
		\centering
		\includegraphics[height=6.5cm]{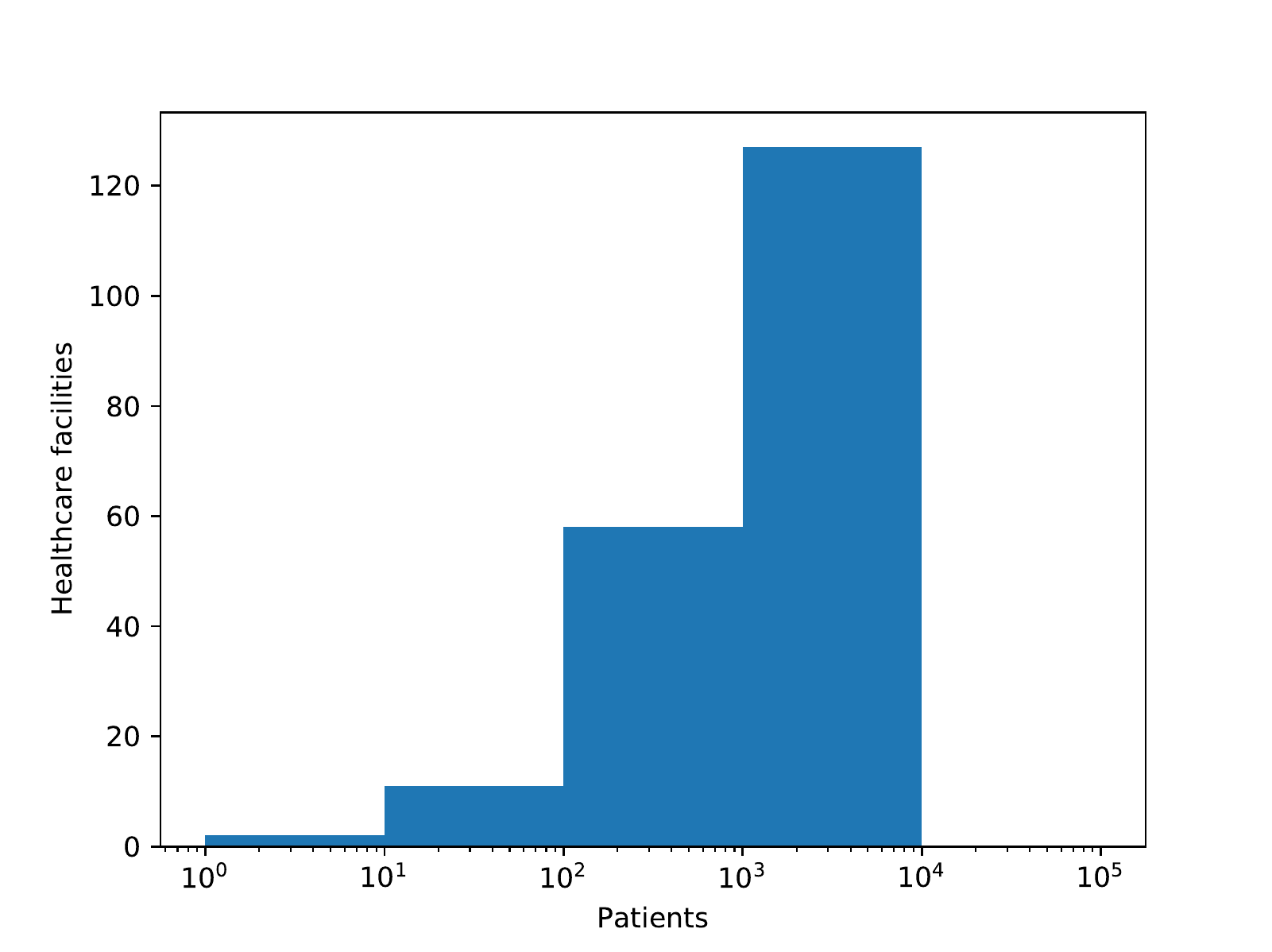}
		\caption{}
	\end{subfigure}  
	~
	\begin{subfigure}[t]{0.5\textwidth}
		\centering
		\includegraphics[height=6.5cm]{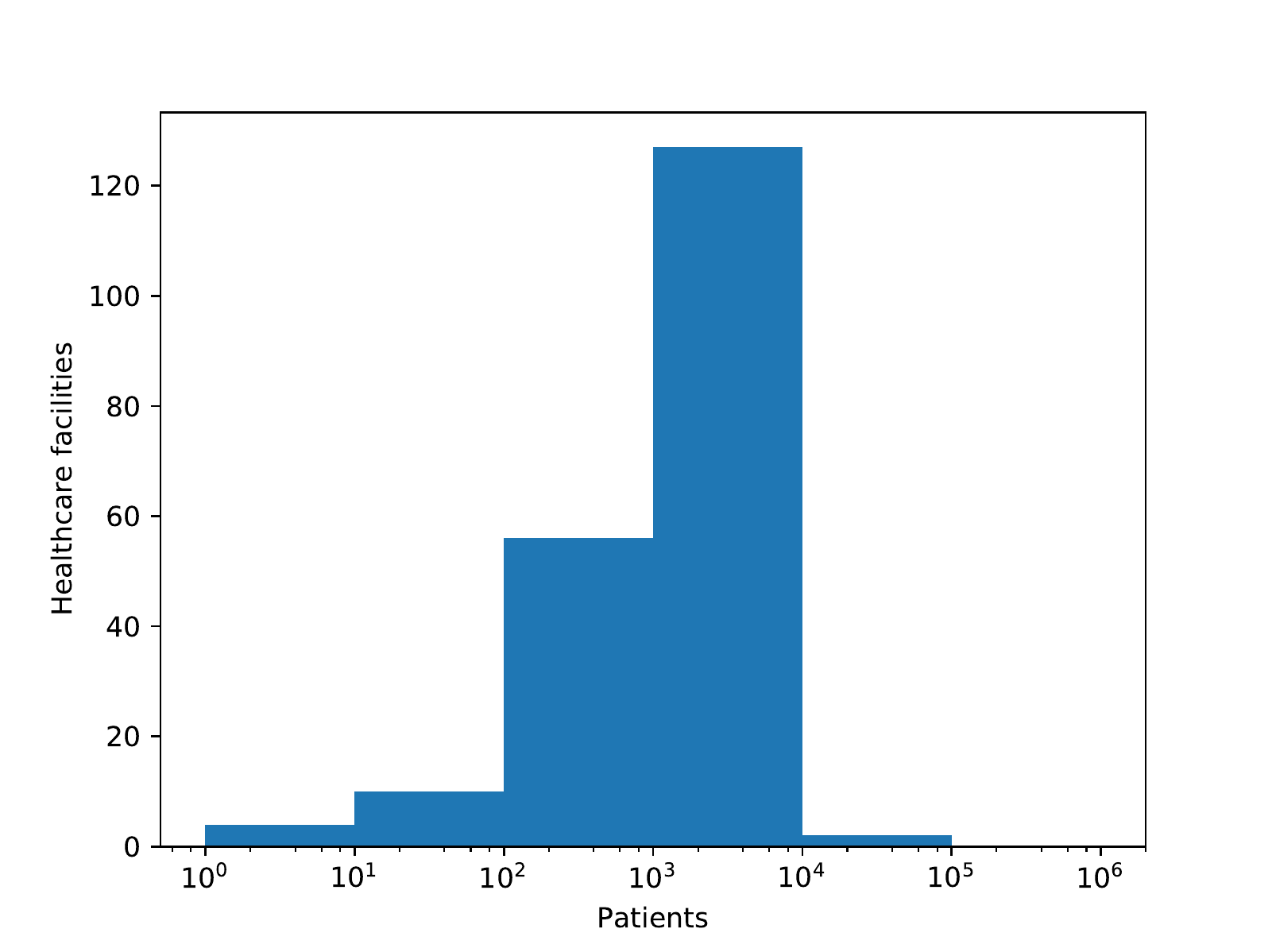}
		\caption{}
	\end{subfigure}%
	~ 
	\begin{subfigure}[t]{0.5\textwidth}
		\centering
		\includegraphics[height=6.5cm]{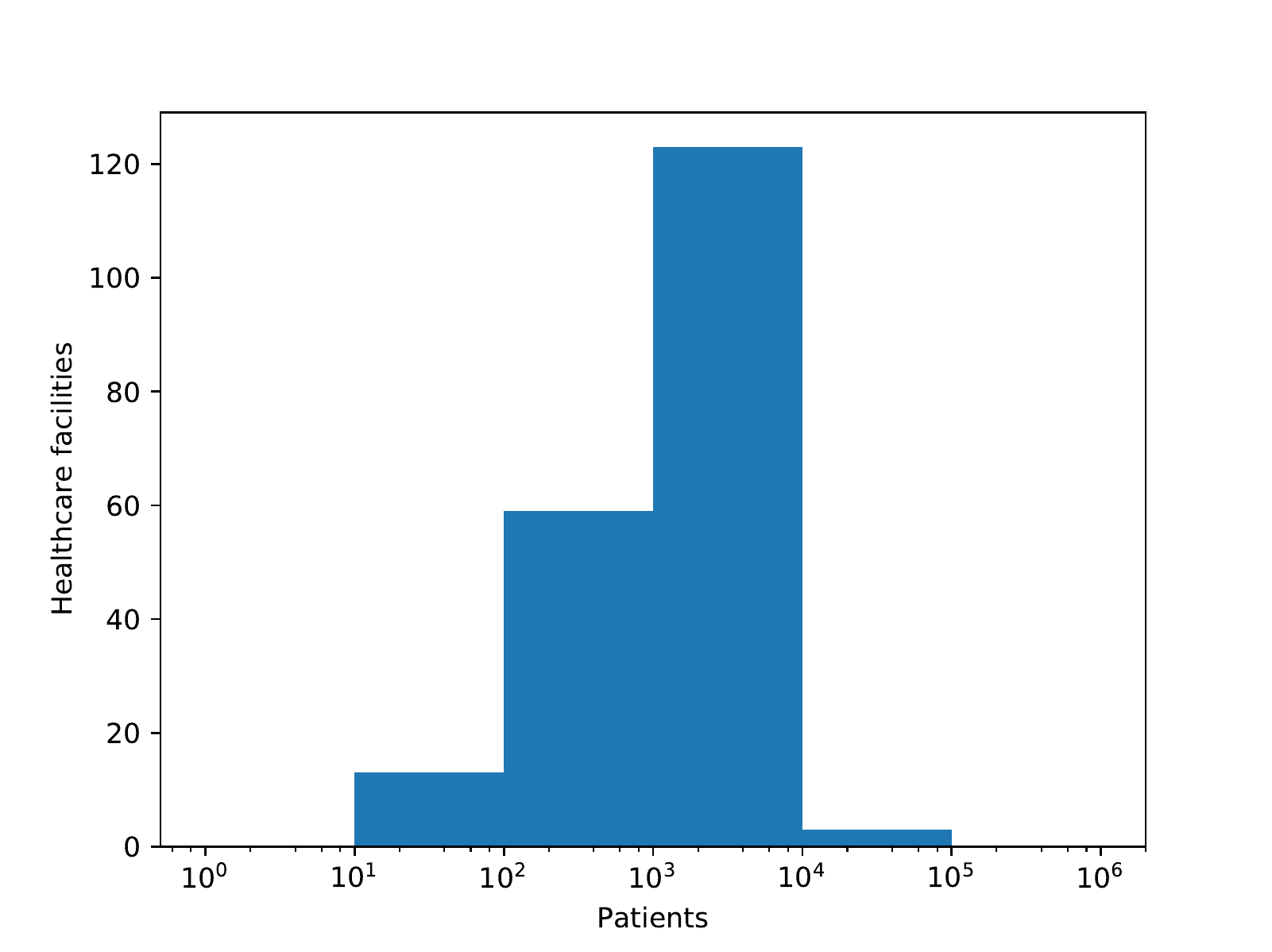}
		\caption{}
	\end{subfigure} 
	~
	\begin{subfigure}[t]{0.5\textwidth}
	\centering
	\includegraphics[height=6.5cm]{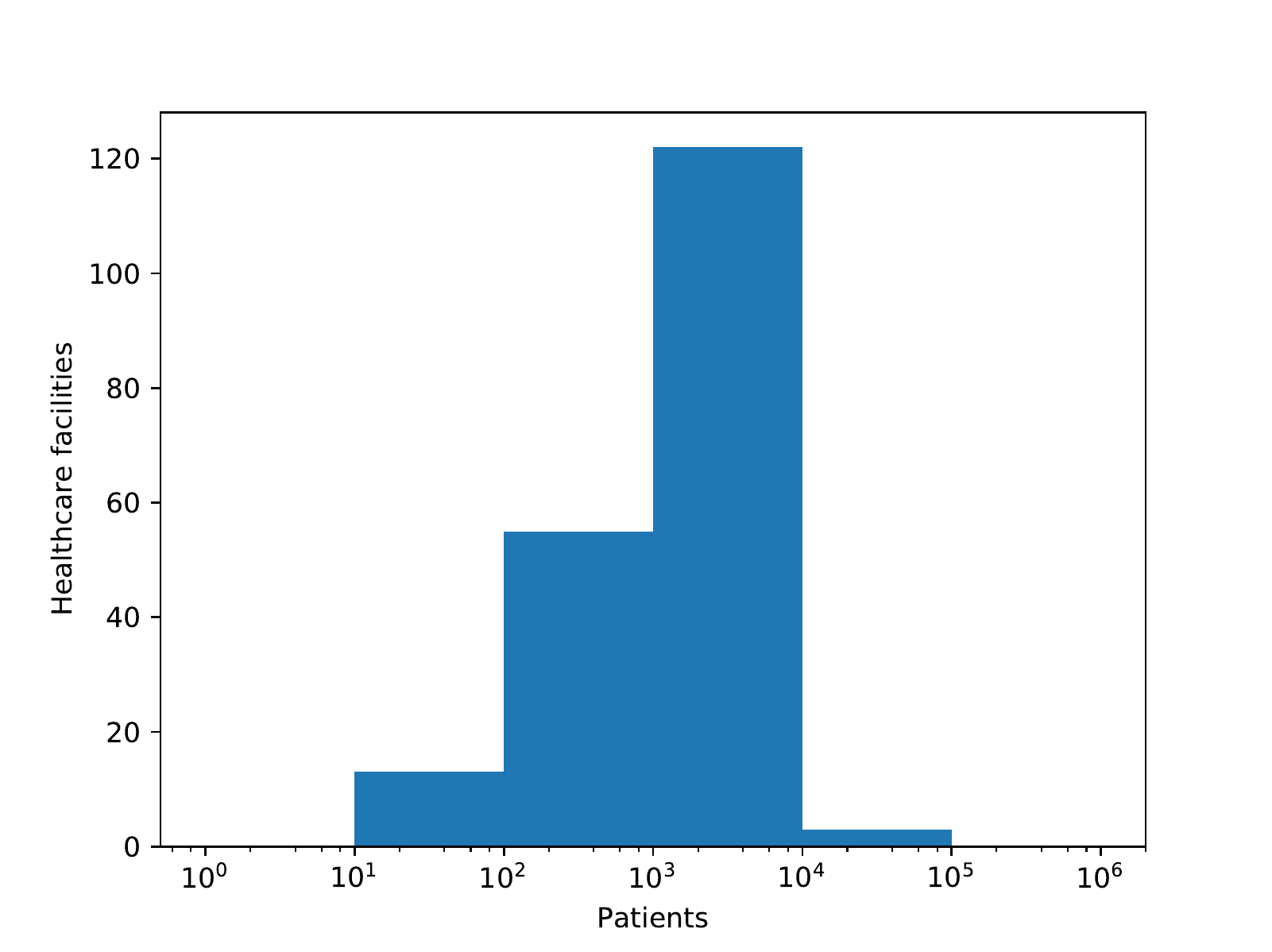}
	\caption{}
	\end{subfigure}%
	~ 
	\begin{subfigure}[t]{0.5\textwidth}
	\centering
	\includegraphics[height=6.5cm]{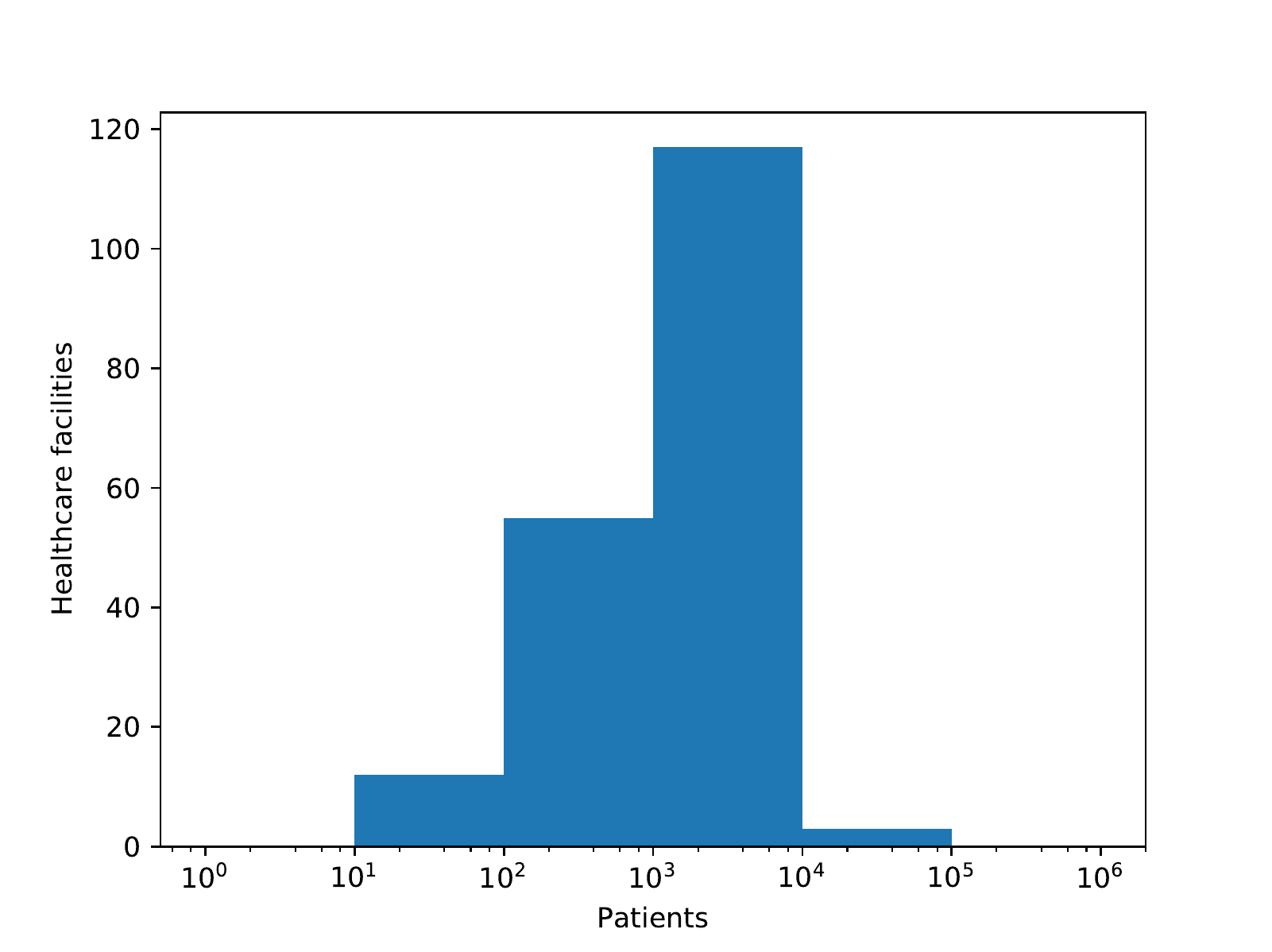}
	\caption{}
	\end{subfigure} 
	\caption{Numbers of healthcare facilities in Lower Saxony having given number of patients reported: (a) within years 2008-2015, (b)~in 2008, (c)~in 2010, (d)~in 2012, (e)~in 2014 and (f)~in 2015.\label{fig:hosp:patients:hkbula03}}
\end{figure}

On the other hand, if we look on the number of patients admitted to the hospitals, compare Figure~\ref{fig:hosp:patients:all} with Figure~\ref{fig:hosp:patients:hkbula03}, we see that for healthcare facilities in all federal states the majority of hospitals had from one up to 10 patients in considered years and if we consider all years at once (Figure~\ref{fig:hosp:patients:all} (a)) most of hospitals registered within year 2008-2015 had from 10 to 99 patients.  
However, if we consider Lower Saxony healthcare facilities collectively the fraction of the hospitals having between 1\,000 and 9\,999 patients dominates.

\begin{figure}
	\centering
	\begin{subfigure}[t]{0.5\textwidth}
		\centering
		\includegraphics[height=6.5cm]{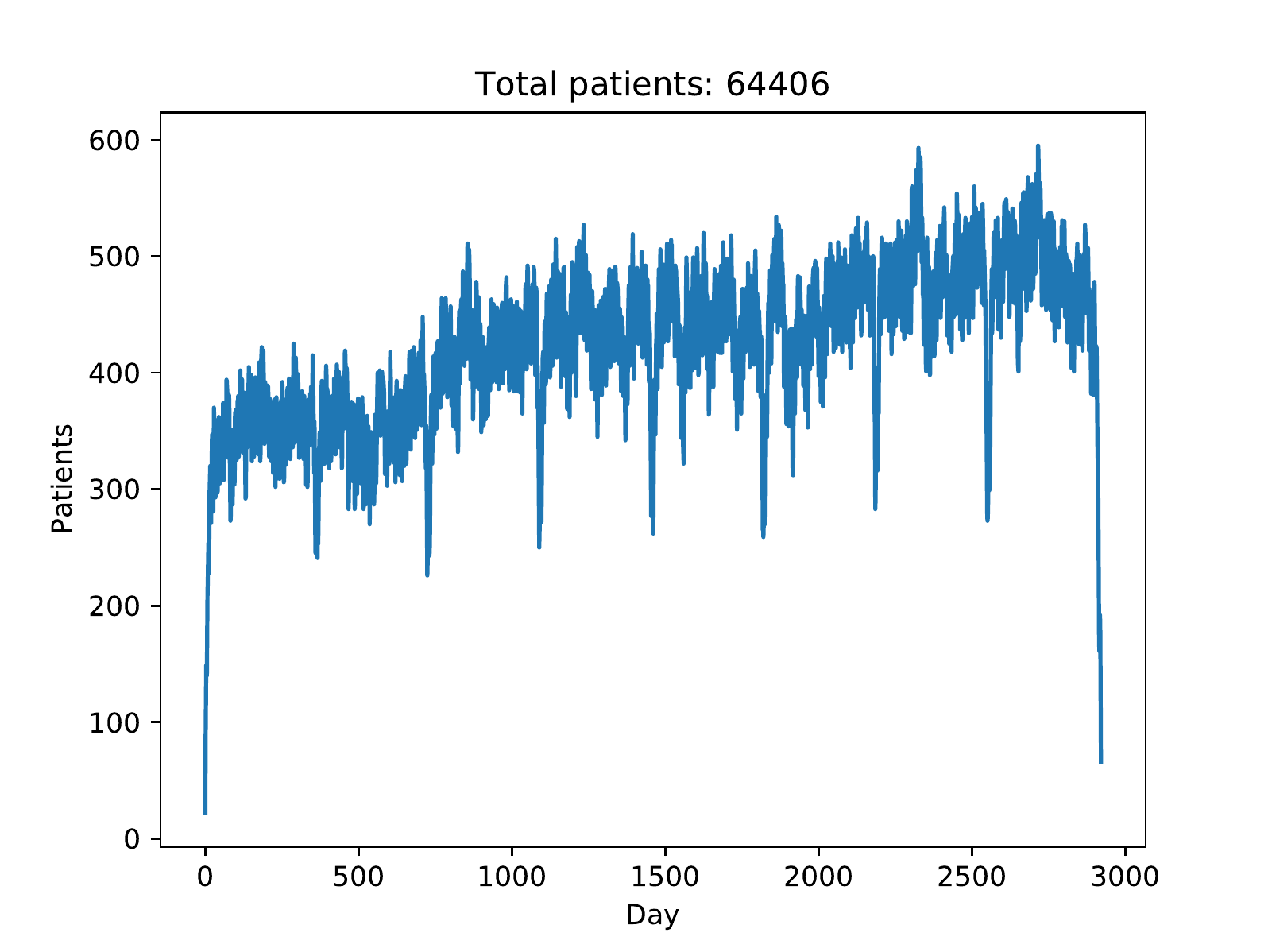}
		\caption{}
	\end{subfigure}%
	~ 
	\begin{subfigure}[t]{0.5\textwidth}
		\centering
		\includegraphics[height=6.5cm]{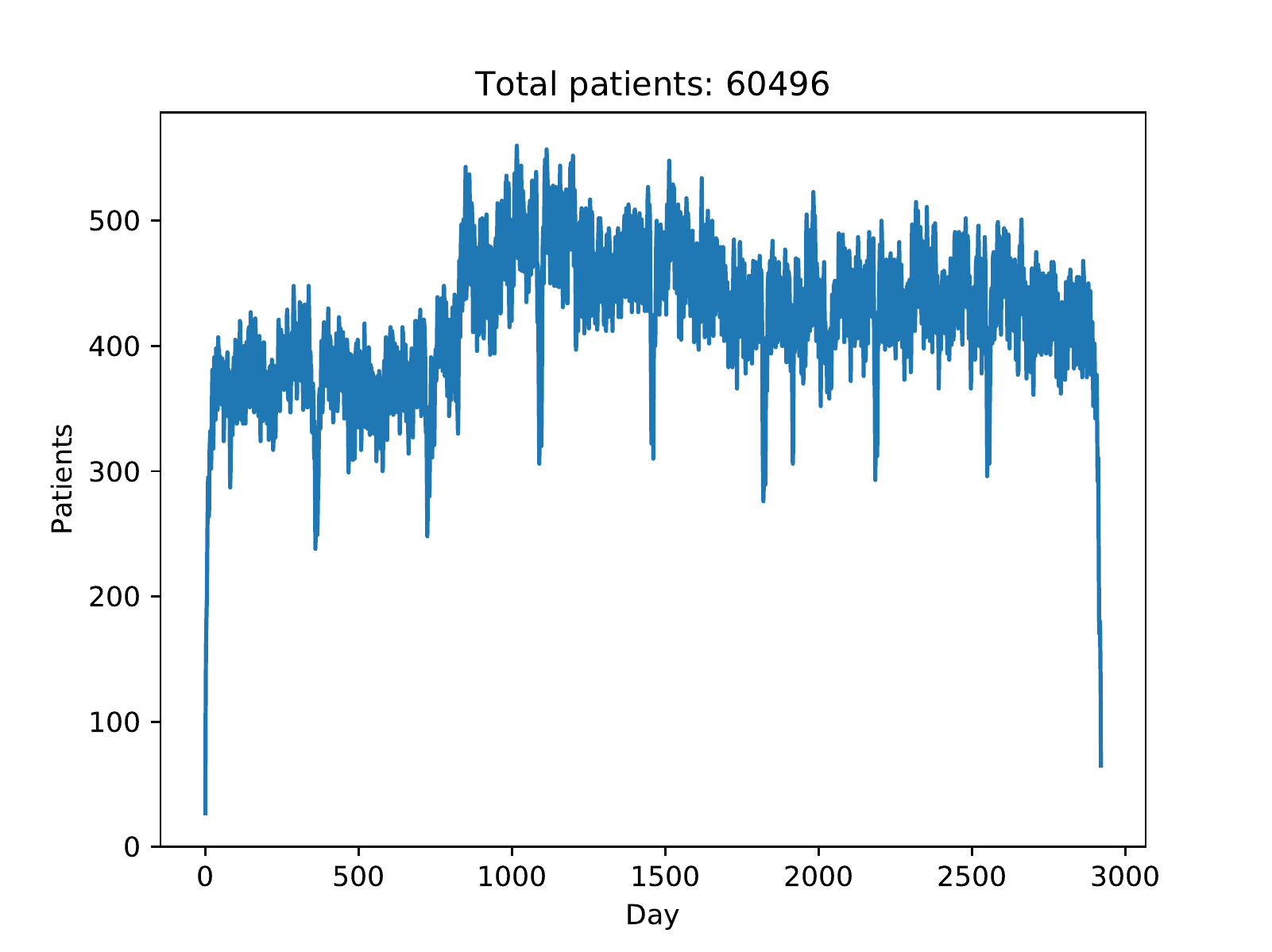}
		\caption{}
	\end{subfigure}  
	~
	\begin{subfigure}[t]{0.5\textwidth}
		\centering
		\includegraphics[height=6.5cm]{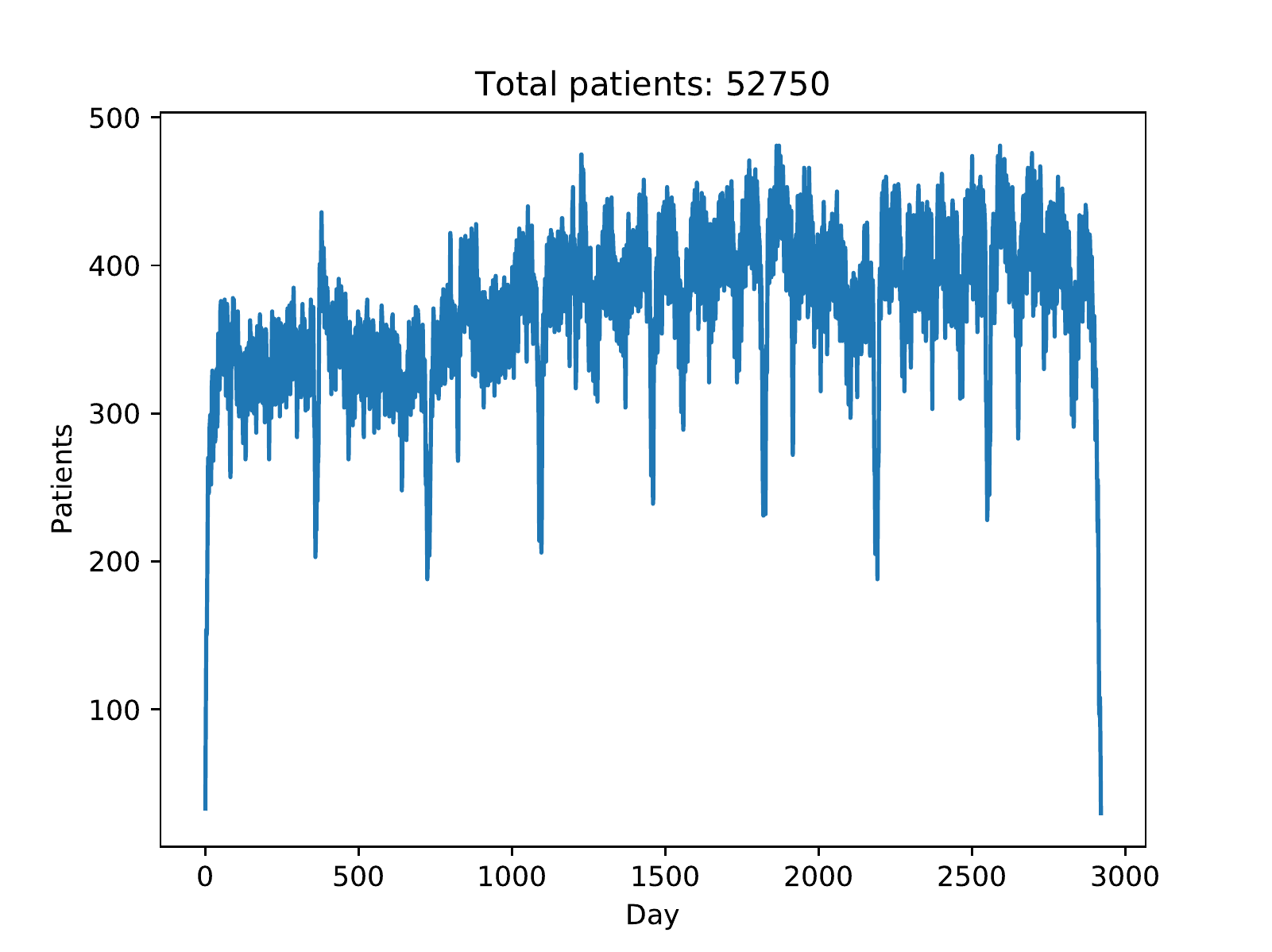}
		\caption{}
	\end{subfigure}%
	~ 
	\begin{subfigure}[t]{0.5\textwidth}
		\centering
		\includegraphics[height=6.5cm]{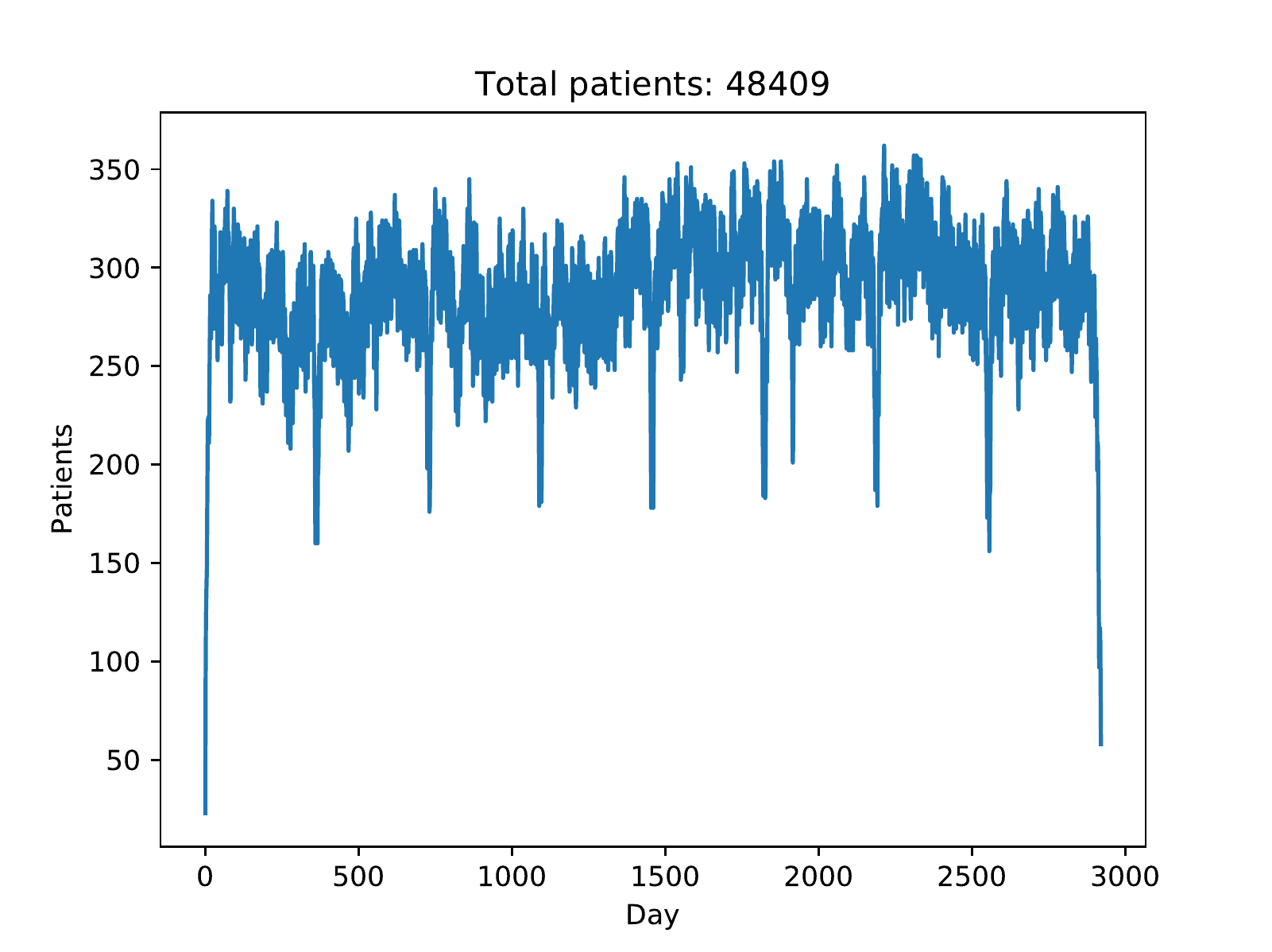}
		\caption{}
	\end{subfigure} 
	~
\begin{subfigure}[t]{0.5\textwidth}
	\centering
	\includegraphics[height=6.5cm]{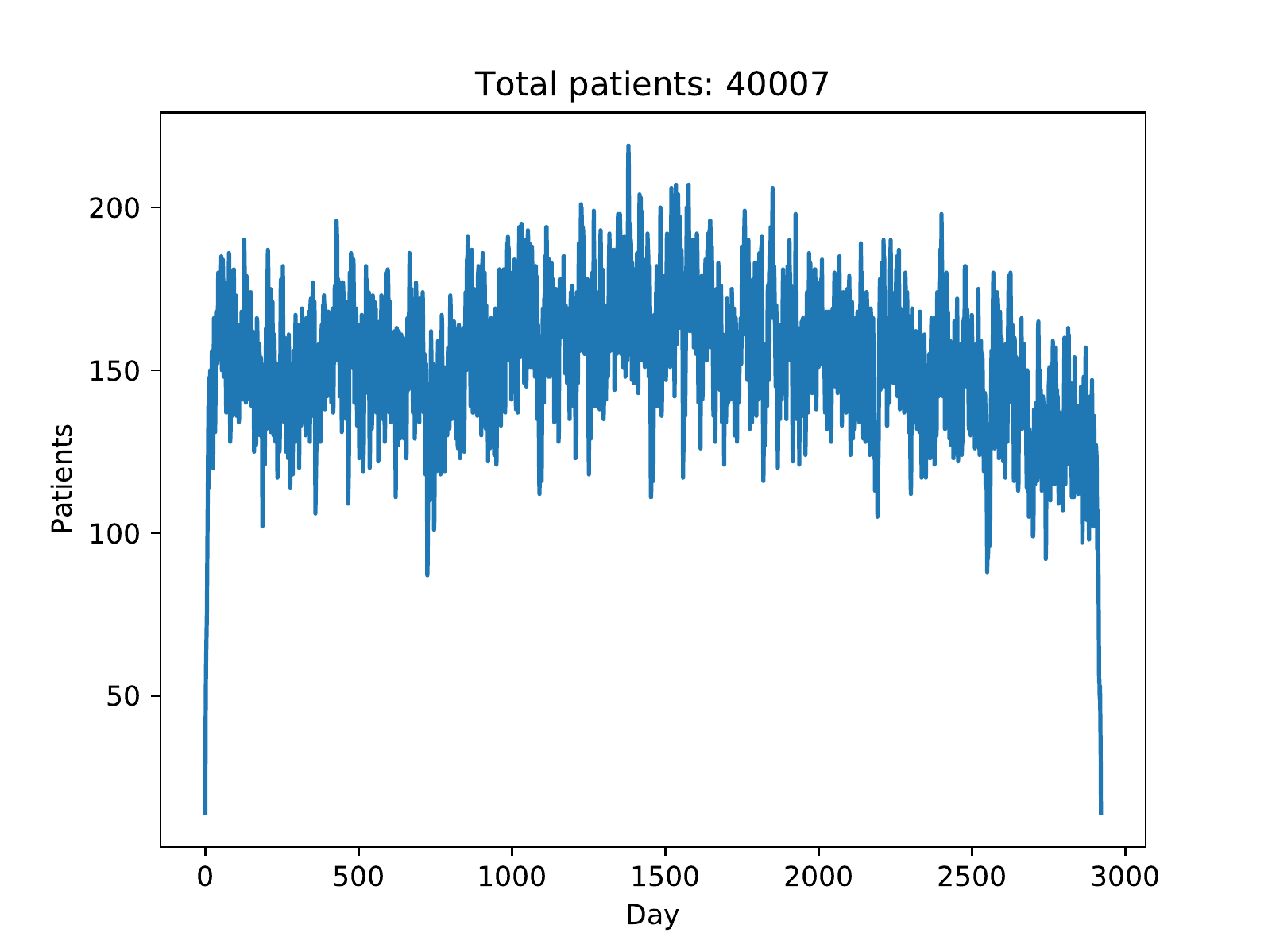}
	\caption{}
\end{subfigure}%
~ 
\begin{subfigure}[t]{0.5\textwidth}
	\centering
	\includegraphics[height=6.5cm]{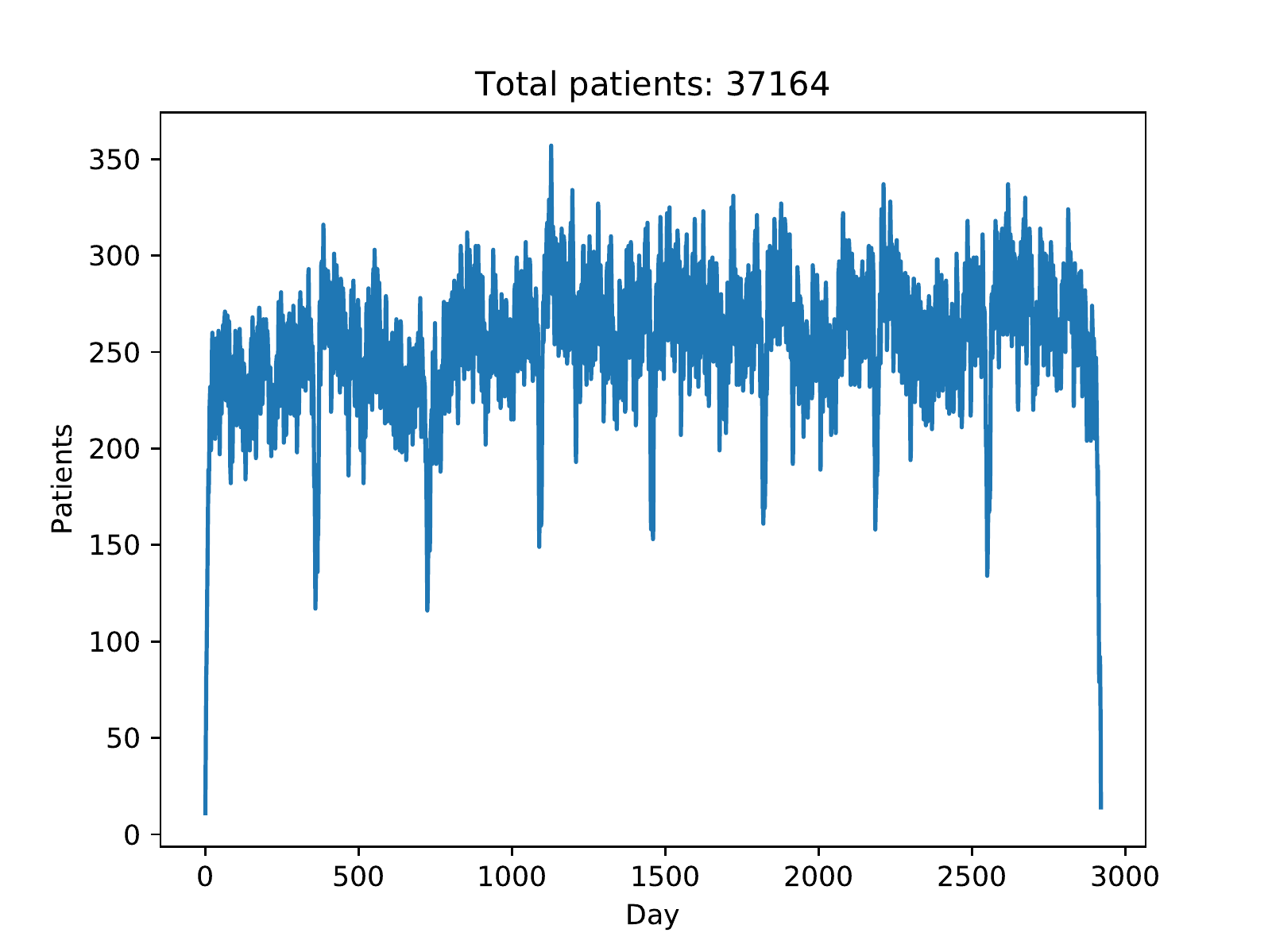}
	\caption{}
\end{subfigure} 
	\caption{Numbers of patients saying in six biggest hospitals in Lower Saxony in time within years 2008-2015. \label{fig:hosp:patients:time}}
\end{figure}

To investigate the number of patients staying in the particular healthcare facilities more deeply we investigative the changes of the number of patients in time (from 2008 up to 2015) for all considered in Lower Saxony healthcare facilities. In Figure~\ref{fig:hosp:patients:time} we presents results for six biggest hospitals. Clearly, we observe some fluctuations and larger deviations in time of the number of patients, but that is rather natural phenomenon where larger deviations can be explained by Christmas time.


\subsection{Durations of stays}

\begin{figure}
\begin{subfigure}[t]{0.5\textwidth}
	\centering
		\includegraphics[height=6.5cm]{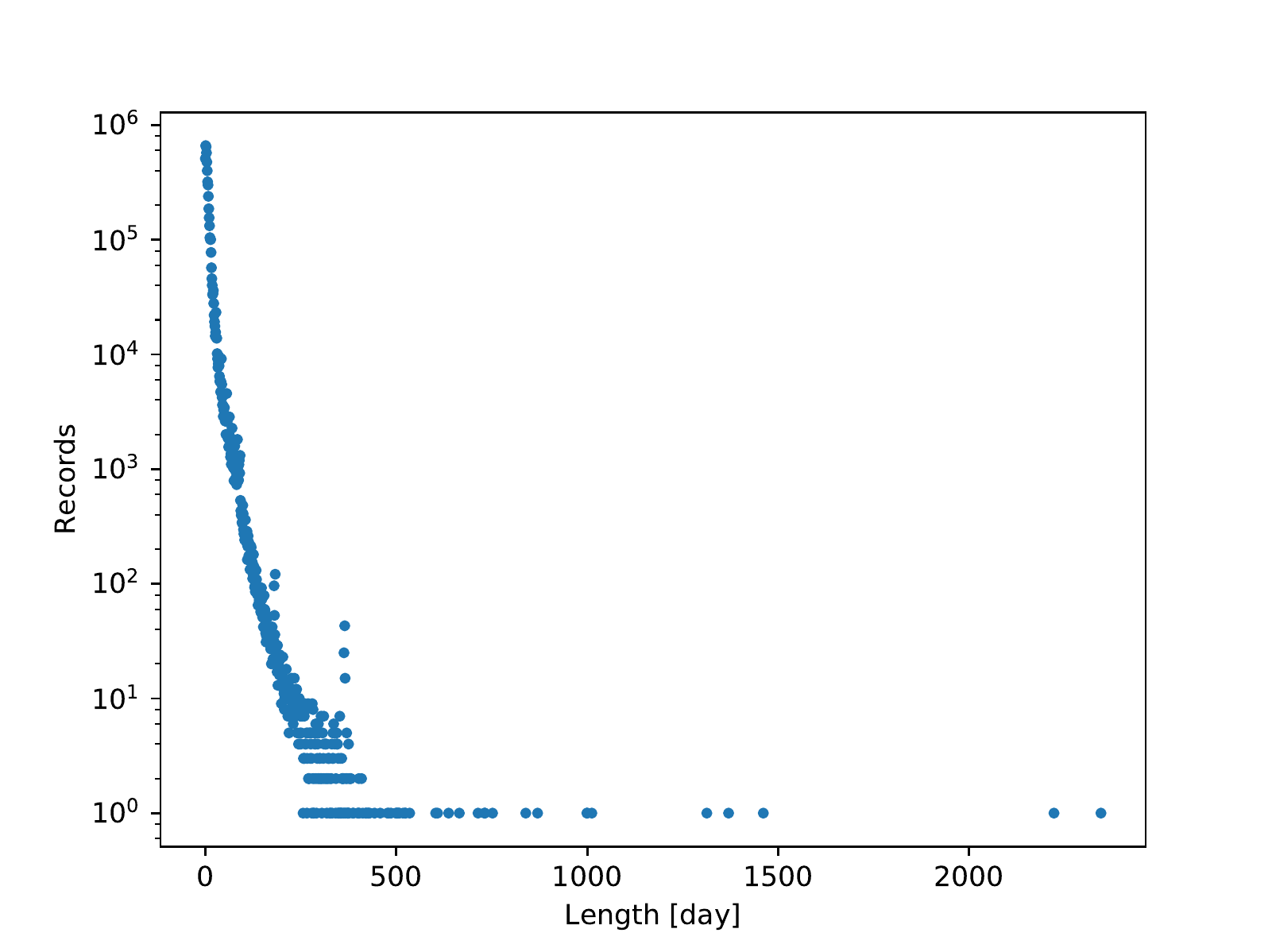}
	\caption{}
\end{subfigure}%
~ 
\begin{subfigure}[t]{0.5\textwidth}
	\centering
		\includegraphics[height=6.5cm]{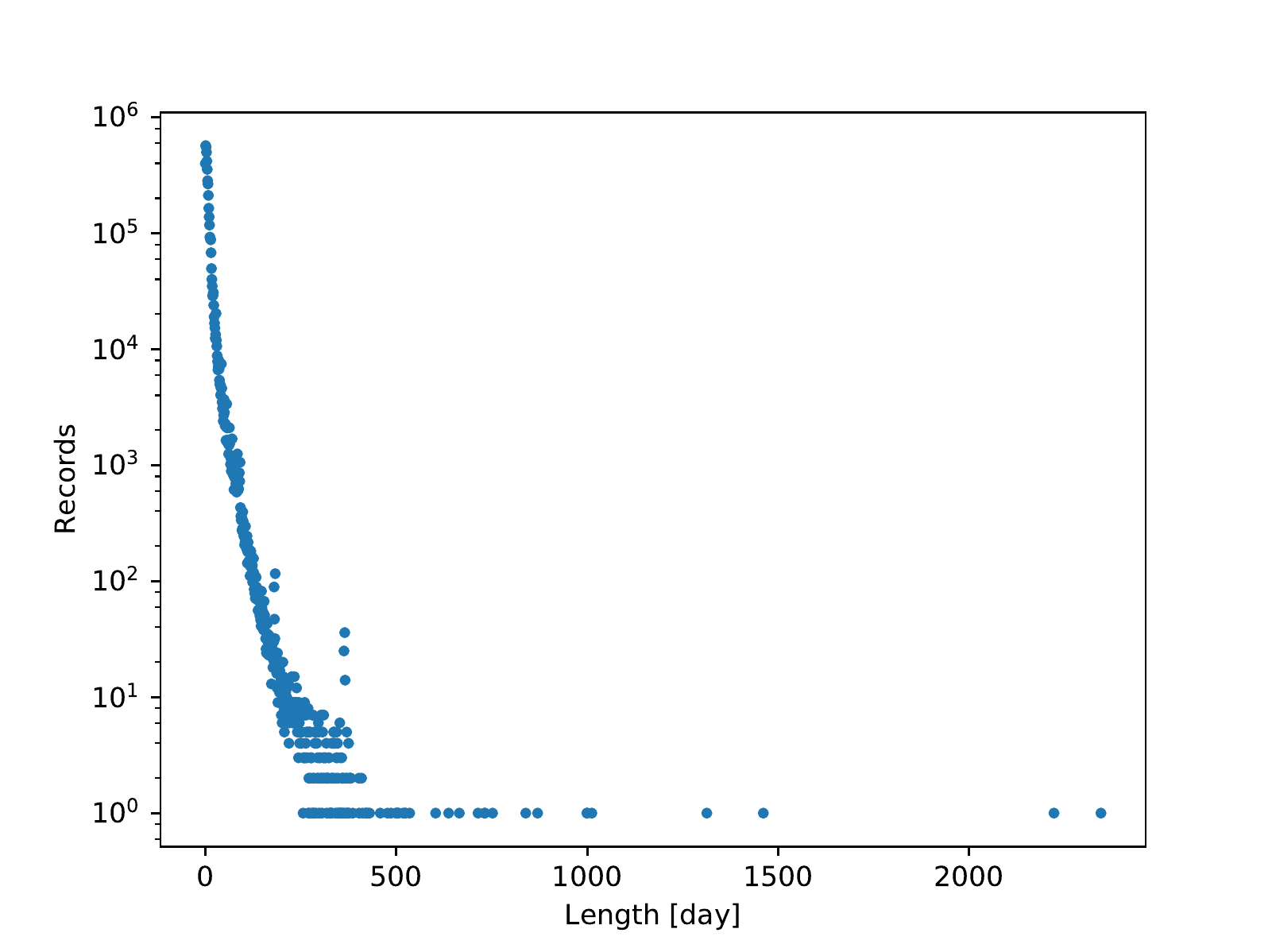}
	\caption{}
\end{subfigure}
\caption{Durations of patients stays in (a) all healthcare facilities, (b) healthcare facilities located in Lower Saxony within years 2008-2015. \label{fig:hosp:stay}}
\end{figure}

We also investigated duration of reported stays of patients in particular healthcare facilities. In Figure~\ref{fig:hosp:stay} we present a histogram of the duration of the hospitalisations for all healthcare facilities (a) and those for Lower Saxony only (b). 

\begin{figure}
	\begin{subfigure}[t]{0.5\textwidth}
		\centering
		\includegraphics[height=6.5cm]{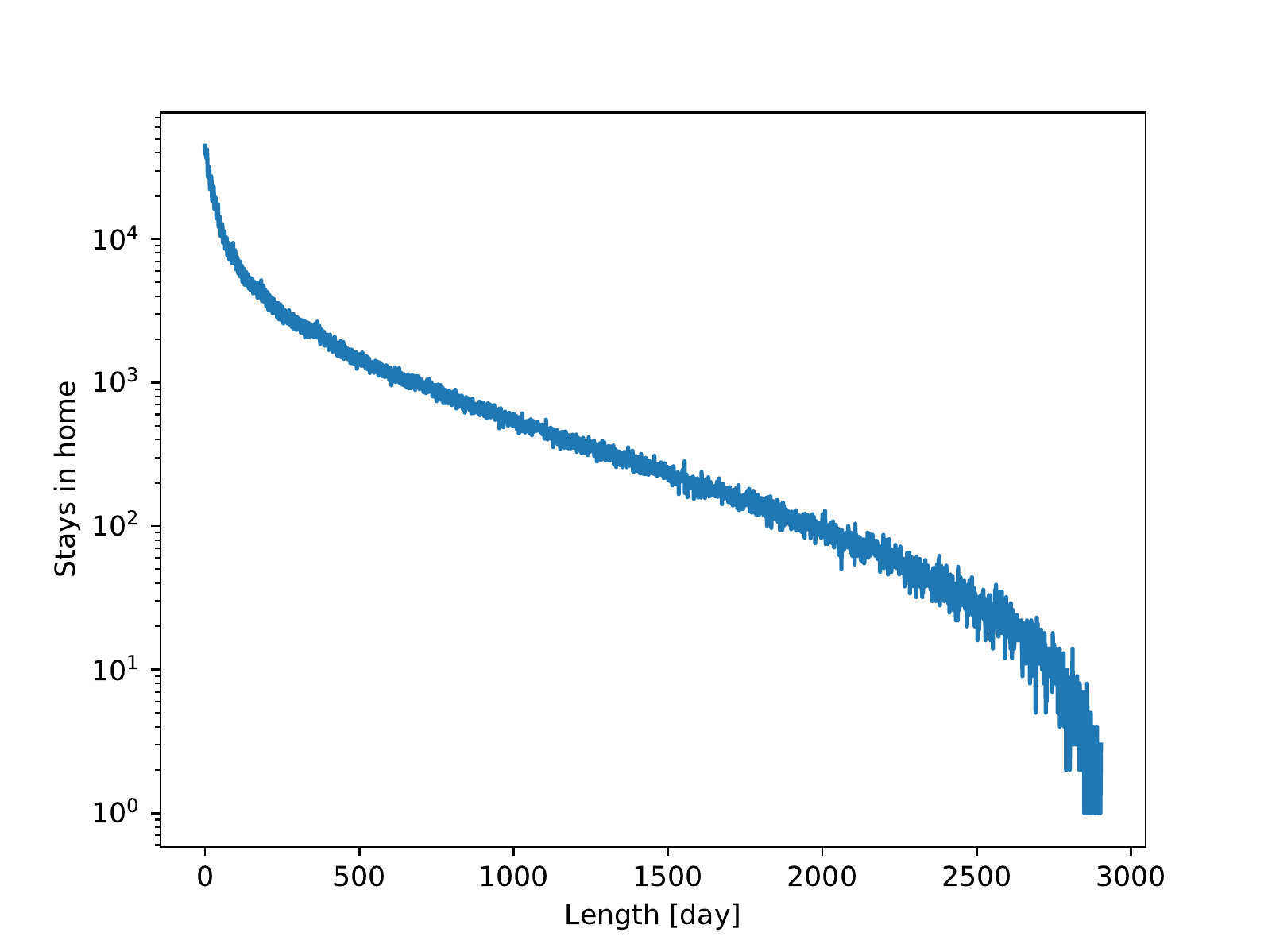}
		\caption{}
	\end{subfigure}%
	~ 
	\begin{subfigure}[t]{0.5\textwidth}
		\centering
		\includegraphics[height=6.5cm]{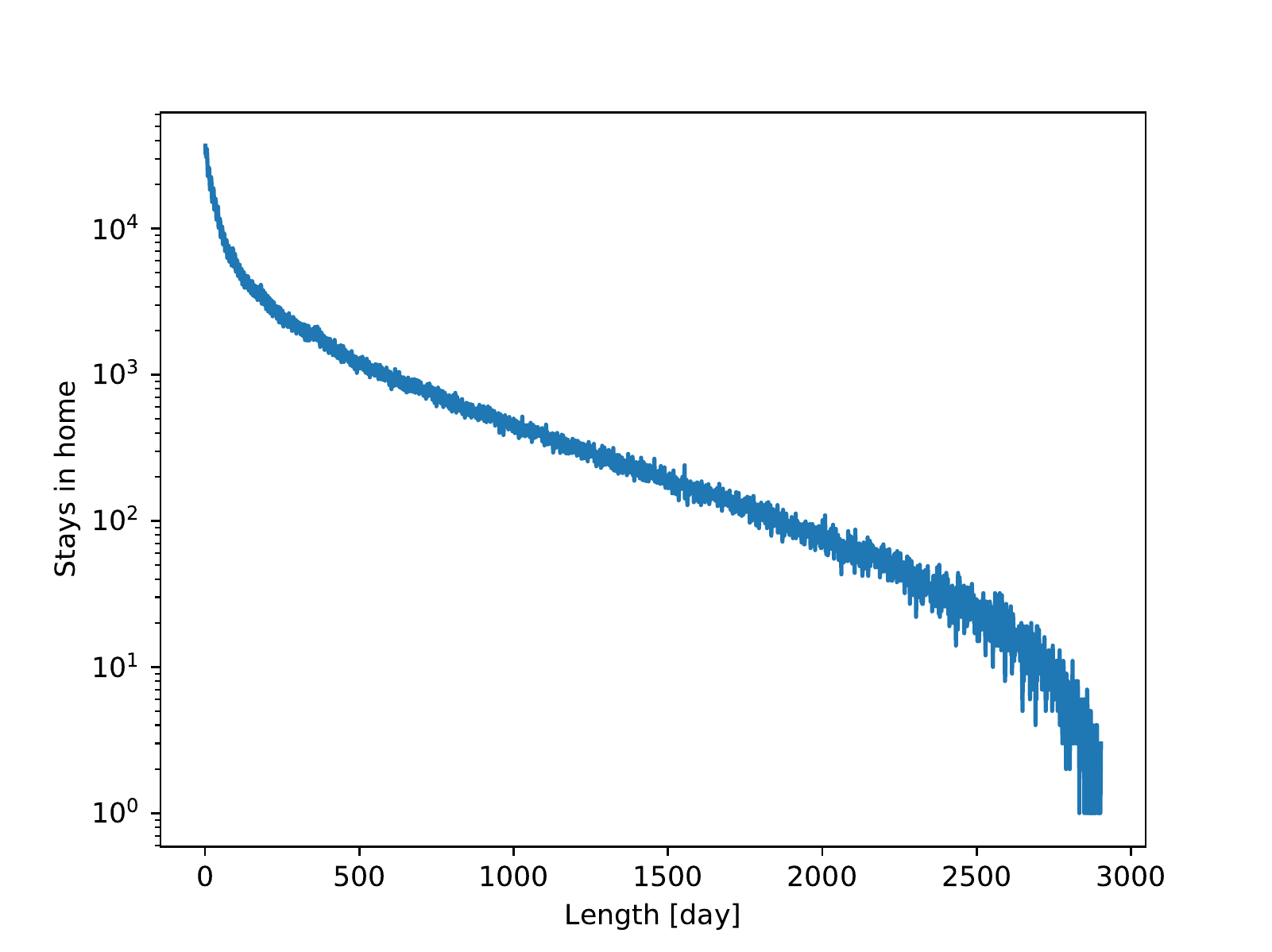}
		\caption{}
	\end{subfigure}

	\caption{Durations of patients stays in society for (a) all healthcare facilities, (b) healthcare facilities located in Lower Saxony within years 2008-2015. \label{fig:hosp:stay:soc}}
\end{figure}

The interesting issue is also duration of the patients stays outside healthcare facilities, understand as a time between hospitalisations spend in society, see Figure~\ref{fig:hosp:stay:soc}, and take that fact into account in the model of the hospital network. 

What is interesting on can not see a big differences between the data for all considered healthcare facilities and those located in Lower Saxony only. 

\subsection{Overlaps}\label{sec:ovelaps}

Analysing provided data for Lower Saxony healthcare facilities we found a number of overlapping records -- 294\,741 cases in total. 
By \emph{overlapping records} we understand distinct sets of two or more records for a given patient, with non-empty intersection of stay periods, either within the same facility or in other facilities.
   
A vast majority of detected overlaps is related to simultaneous stay of some patients in at most two institutions, but in general there are cases when a patient is attributed to five stay records in the same day. 

Performing more detailed
analysis of the data we come up with the following classification:
\begin{itemize}
	\item standard transfer --- one day overlap of two stay periods, where both periods are longer than one day and each record corresponds to different facility,
	\item first day transfer/last day transfer --- similar to above, but duration of the stay in one facility is exactly one day long and it coincides with admission to/discharge from the latter facility,
	\item simultaneous two admissions in a single institution --- two reported stays in the same place for the same period,
	\item temporary transfer  --- two records, period of one of them is contained in the other, and admission and discharge dates are not the same,
	\item simultaneous two entries in two different institutions  --- periods are exactly the same, but the facilities are different,
	\item unknown two admissions in two different institutions --- any two records for hospitalisations in different institutions, which is not covered by the cases already introduced,
	\item two admissions in a single institution --- two reported stays in the same institution but for different (overlapping) periods,
	\item unknown multiple admissions ($n$) --- more than two records of overlapping hospitalisation periods, with maximal number of records in a given day is $n$.
\end{itemize}

In Figure~\ref{fig:examp} we present an exemplary visualization of the overlapping appearing in the data base.

\begin{figure}
{
	\begin{verbatim}
	
	0: | 2015-06-30:   ###### |
	1: | 2015-06-30: ###      |
	
	1: | 2013-02-01:                     # |
	2: | 2013-02-01: ##################### |
	
	0: | 2014-12-03:                #####                   |
	2: | 2014-12-03: ###################################### |
	
	0: | 2013-05-25: ####### |
	1: | 2013-05-25: ###     |
	
	0: | 2013-07-09:                     ####                       |
	1: | 2013-07-09:                        ########                |
	2: | 2013-07-09: ############################################## |
	
	0: | 2015-01-13: ##############  |
	2: | 2015-01-13:          ###### |
	
	0: | 2013-09-06: ######## |
	1: | 2013-09-06: ######## |	
		
	1: | 2013-11-08: ##################### |
	1: | 2013-11-08: ##################### |	
	\end{verbatim}
}\caption{Examples of overlaps. In the first column the healthcare facility number is given, next initial date of any hospitalisation and finally the graphical representation of hospitalisation duration (sign $\#$ denotes one day of stay in the healthcare facility). In the first row we see the example of {\it standard transfer}, next {\it last day transfer}, {\it temporary transfer}, {\it unknown two admissions in two institutions}, {\it unknown multiple admissions (3)},  another example of {\it unknown two admissions in two institutions}, {\it simultaneous two admissions in two institutions}, {\it simultaneous two admissions in a single institution}. \label{fig:examp}}
\end{figure} 

For number of records and the percentage in all the overlaps see~Table\ref{tab:overlaps}. 

In general, these overlapping hospitalisation records allow us to deduce direct transfers between healthcare facilities. Therefore we focus our analysis on the cases, where these transfers can be deduced with high probability of success using the provided hospitalisation periods. We are not interested in detailed analysis of minor problematic cases (unknown multiple entries ($n$)). These are rare and they are often impossible to analyse reliably without study of individual cases.


\begin{table}
	\centering
	\caption{Identified types of overlaps in AOK Lower Saxon data.}\label{tab:overlaps}
\begin{tabular}{|r|r|r|}
	\hline 
{\bf type} &	\multirow{2}{*}{{\bf Overlap description}} & {\bf number of records}\\
     &                                   & {\bf (\% of all detected overlaps)}\\
	\hline
1 & standard transfer & 112\,368 (38.1\%)\\ 	\hline 
2 & two admissions in a single institution &  69\,788 (23.7\%)\\ 	\hline 
3 & simultaneous two admissions in a single institution &  69\,577 (23.6\%)\\ 	\hline 
4 & first day transfer &  23\,139 ( 7.9\%)\\ 	\hline 
5 & temporary transfer &  11\,992 ( 4.1\%)\\ 	\hline 
6 & unknown two admissions in two institutions &   3\,645 ( 1.2\%)\\ 	\hline 
7 & unknown multiple admissions (3) &   2\,402 ( 0.8\%)\\ 	\hline 
8 & last day transfer &    1\,208 ( 0.4\%)\\ 	\hline 
9 & simultaneous two admissions in two institutions &    606  ( 0.2\%)\\ 	\hline 
10 & unknown multiple admissions (4+) &     16 ( 0.0\%)\\ 	\hline 
\end{tabular} 
\end{table}

\begin{figure}
	 \centering
	\begin{subfigure}[t]{0.5\textwidth}
		\centering
		\includegraphics[height=6.5cm]{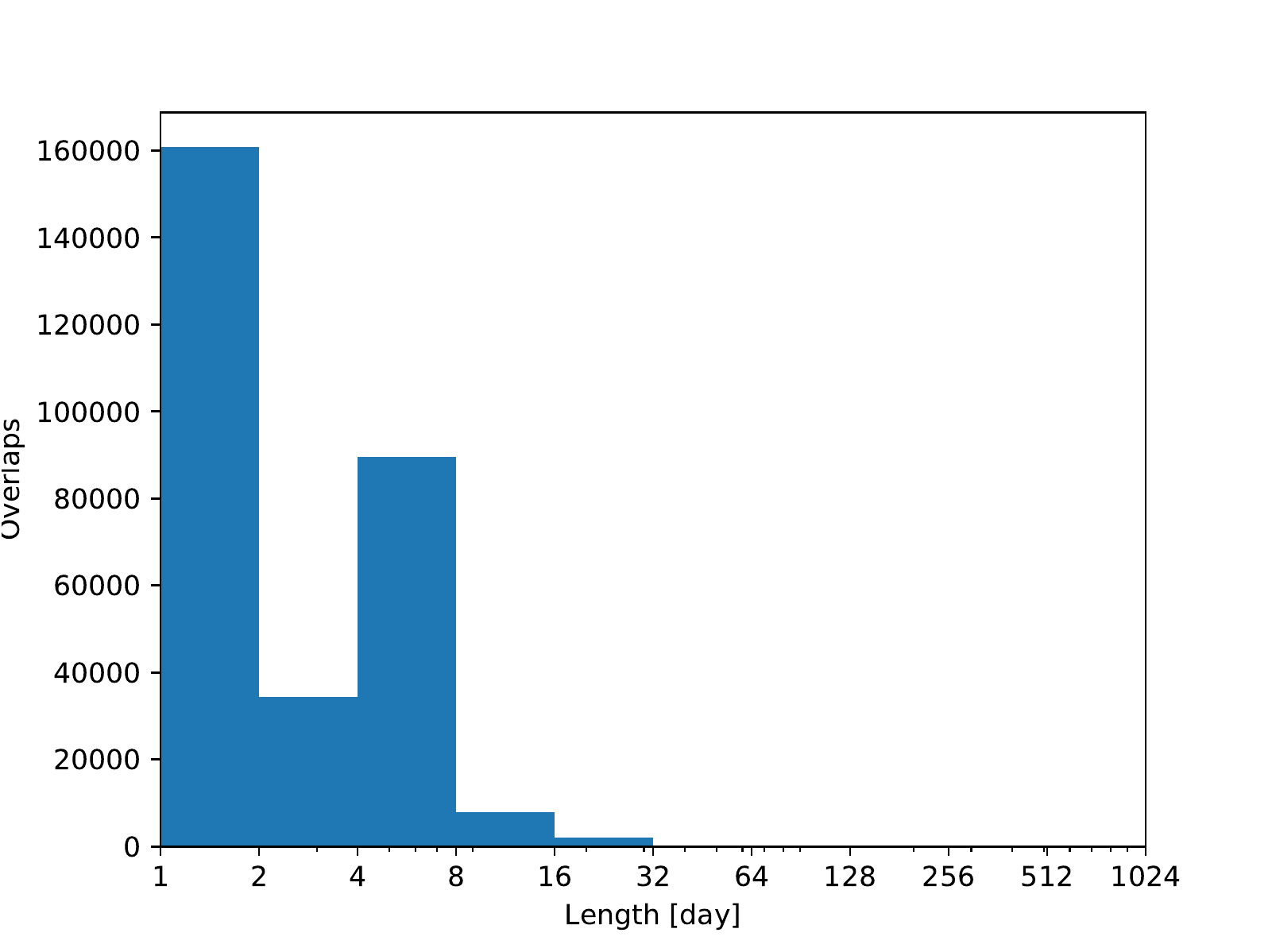}
		\caption{}
	\end{subfigure}%
	~ 
	\begin{subfigure}[t]{0.5\textwidth}
		\centering
		\includegraphics[height=6.5cm]{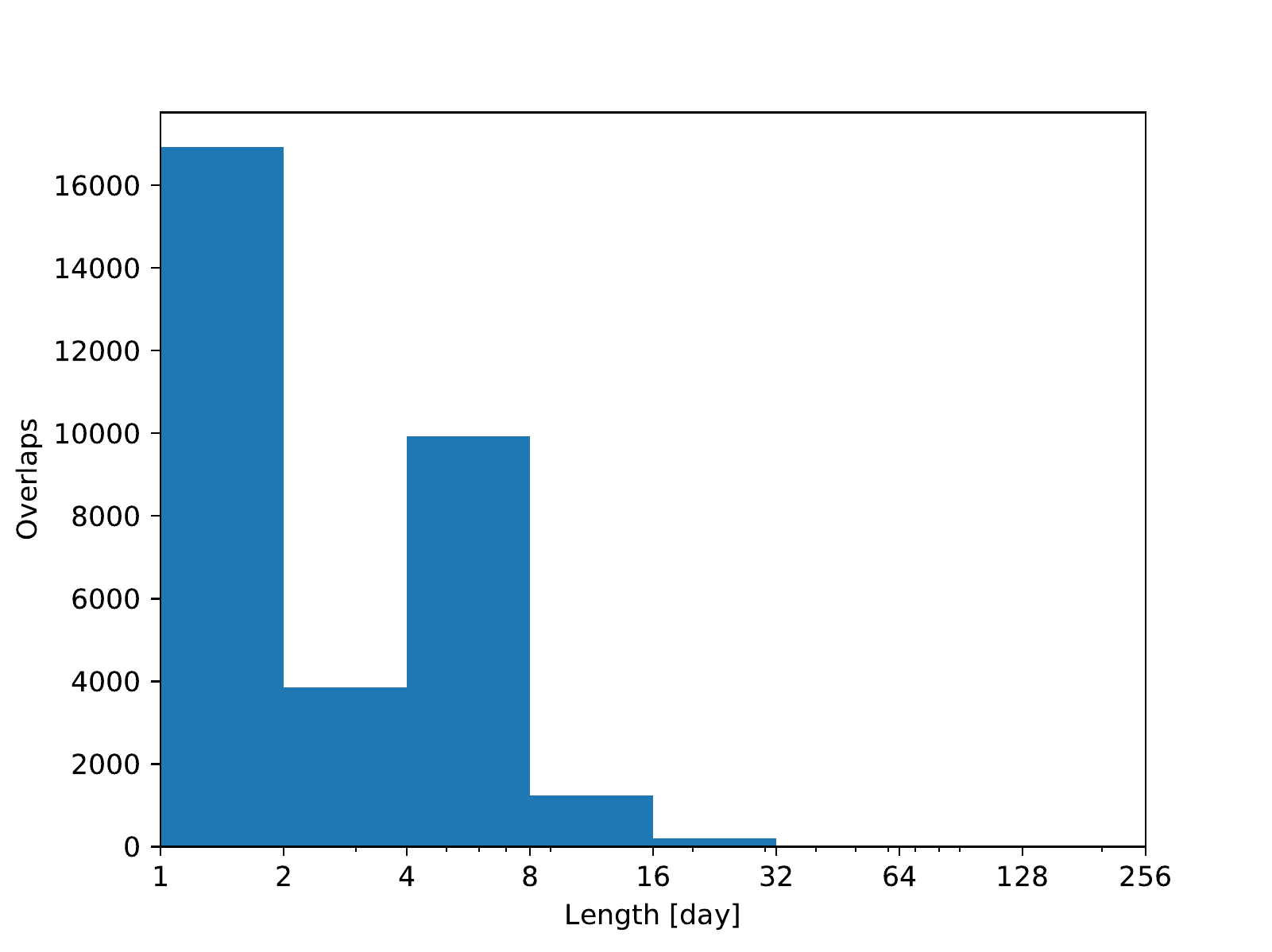}
		\caption{}
	\end{subfigure}  
	~
		 \centering
	\begin{subfigure}[t]{0.5\textwidth}
		\centering
		\includegraphics[height=6.5cm]{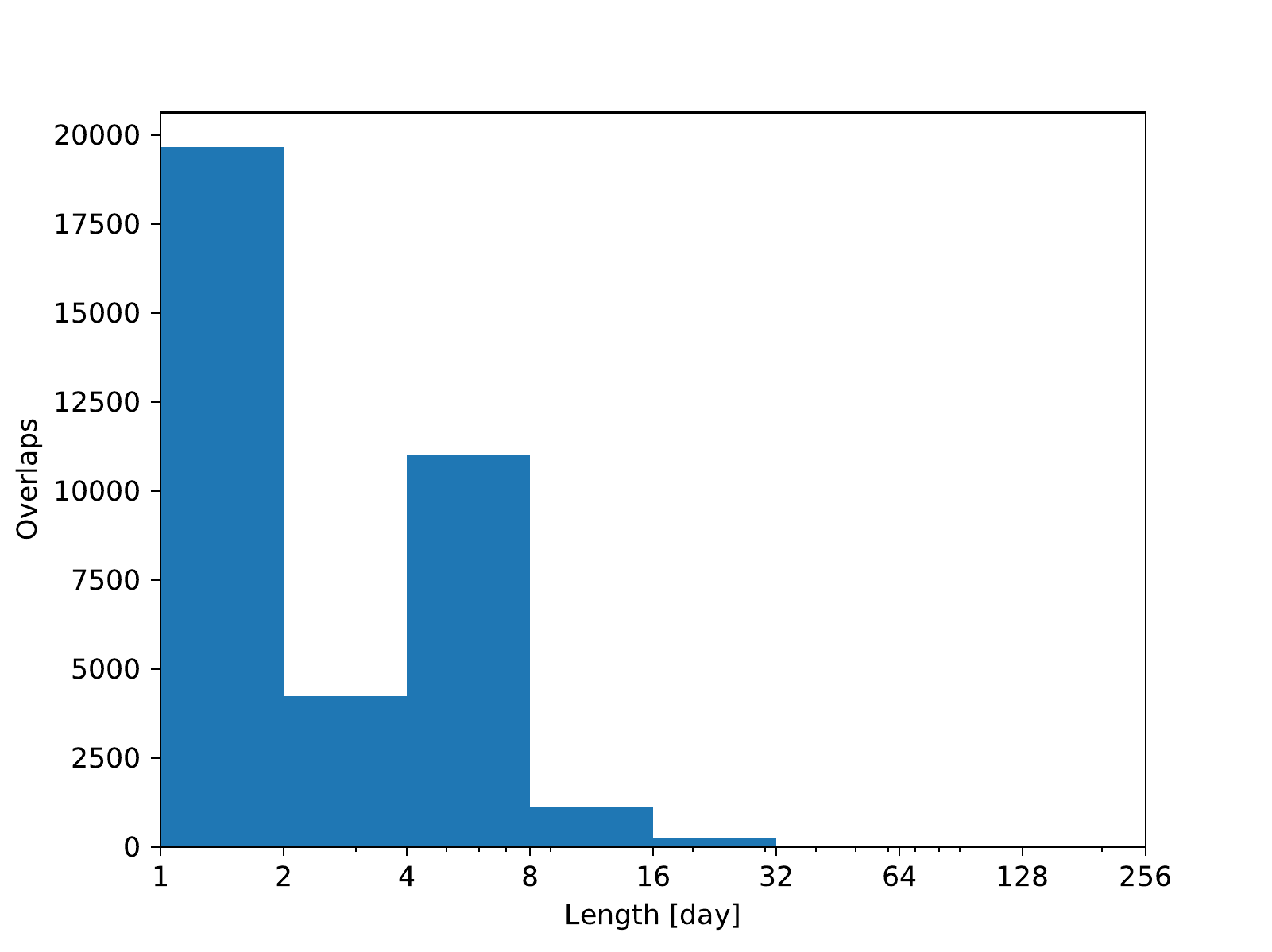}
		\caption{}
	\end{subfigure}%
	~ 
	\begin{subfigure}[t]{0.5\textwidth}
		\centering
		\includegraphics[height=6.5cm]{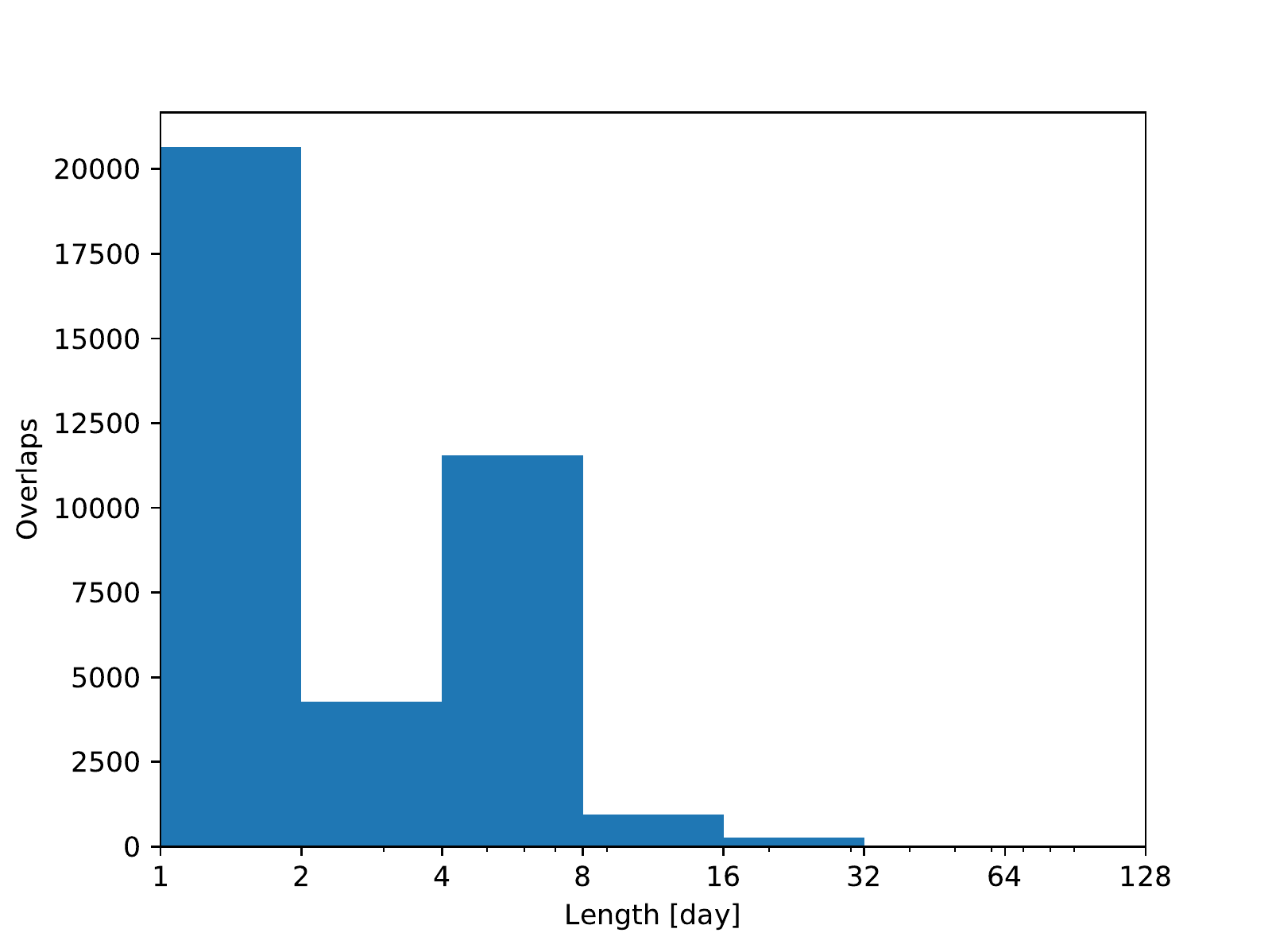}
		\caption{}
	\end{subfigure}  
	~
	\begin{subfigure}[t]{0.5\textwidth}
	\centering
	\includegraphics[height=6.5cm]{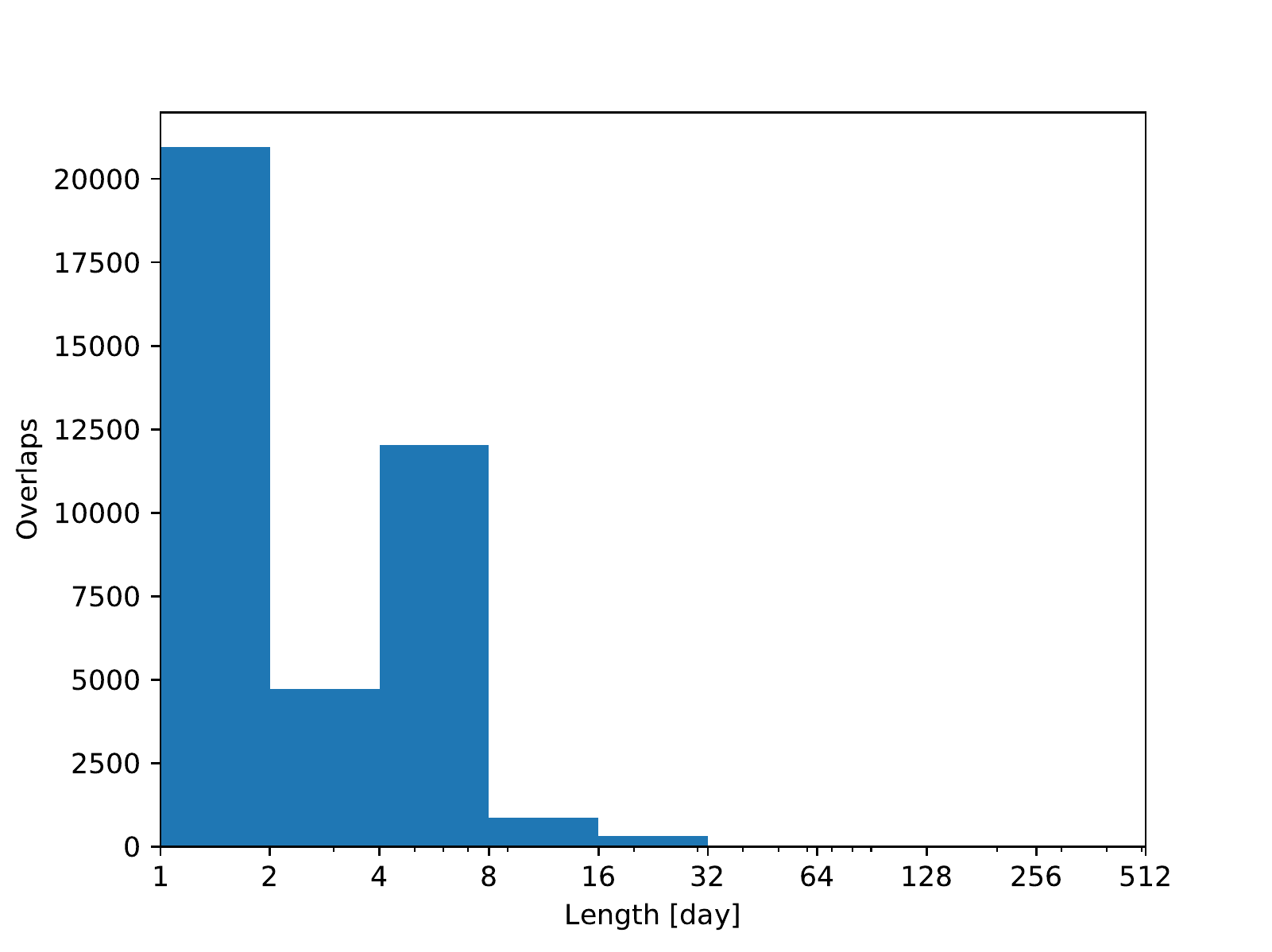}
	\caption{}
	\end{subfigure}%
	~ 
	\begin{subfigure}[t]{0.5\textwidth}
	\centering
	\includegraphics[height=6.5cm]{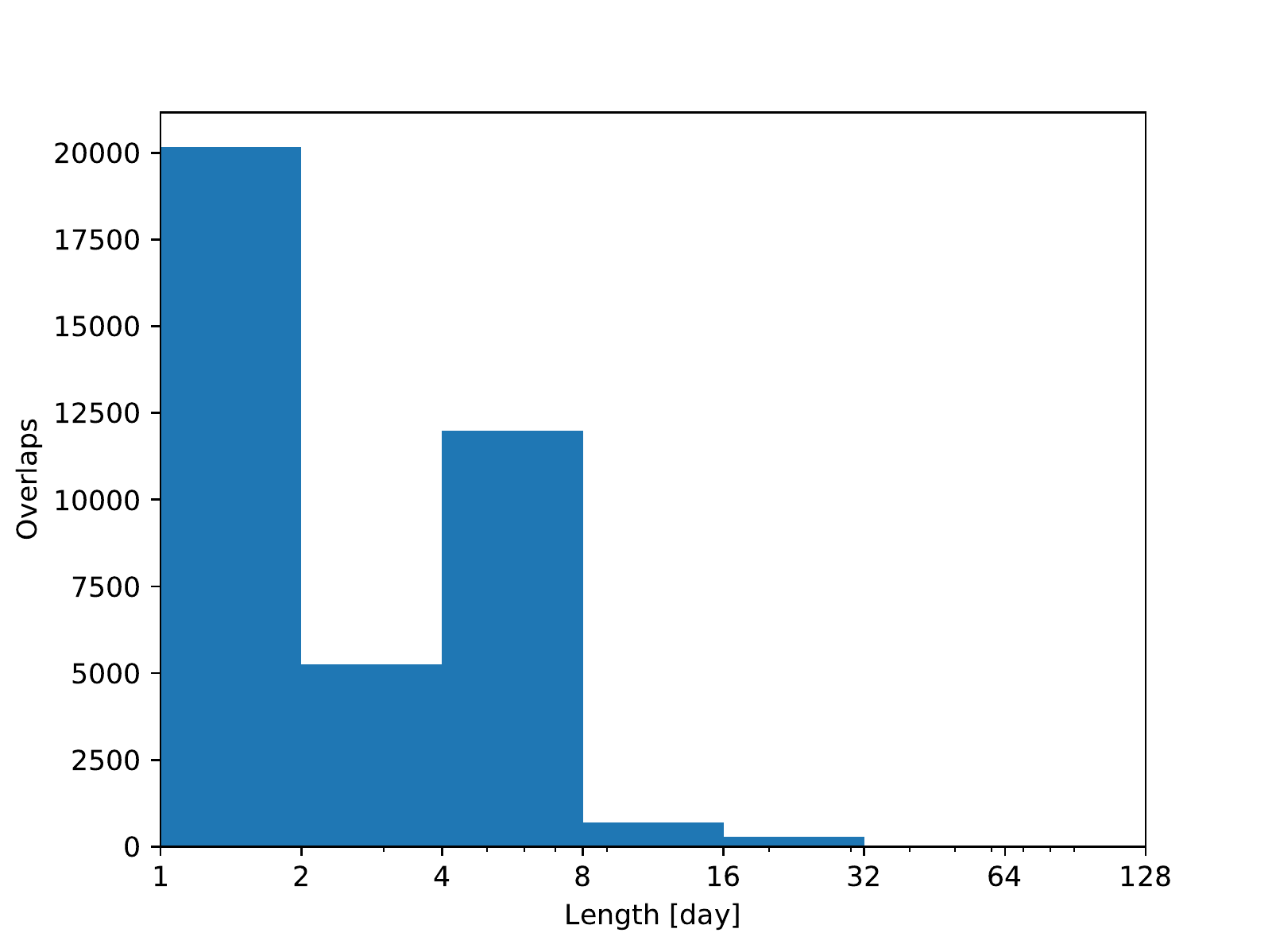}
	\caption{}
	\end{subfigure} 
	\caption{Number of detected overlaps as a function of patient's duration of stays for Lower Saxony healthcare facilities reported: (a) within years 2008-2015, (b) in 2008, (c) in 2010, (d) in 2012, (e) in 2014 and (f) in 2015.\label{fig:czas:overlap:hkbula03}}
\end{figure}

In general over 54\% (160\,764) of overlapping periods within years 2008--2015 intersect by one day,  about 14\% of them are 4-day overlaps (42\,763), while both 3 (22\,913) and 5-day overlapping records (25\,409) are both around 8\% of all overlaps,
compare with~Figure~\ref{fig:czas:overlap:hkbula03} (a).
Interestingly 2-day overlaps are below 4\% (11\,542), which are less common that 6-day overlaps (over 5\%, 15\,027).

In order to deeply characterize the overlapping types we use a four-digit calcification. 
The truth is indicated by 1 while 0 means false. First digit indicate if two considered  overlaps have place in the same healthcare facility, second digit: if overlaps have the same diagnoses, third: 
if two overlaps have the same admission dates and fourth if two overlaps have the same discharge dates.
For example code 1100 simply means that two considered overlaps have been reported by the same healthcare facility, in both cases the diagnosis was the same, but there were different dates of admissions and discharges.  

\begin{figure}
	\centering
	\includegraphics[height=17cm]{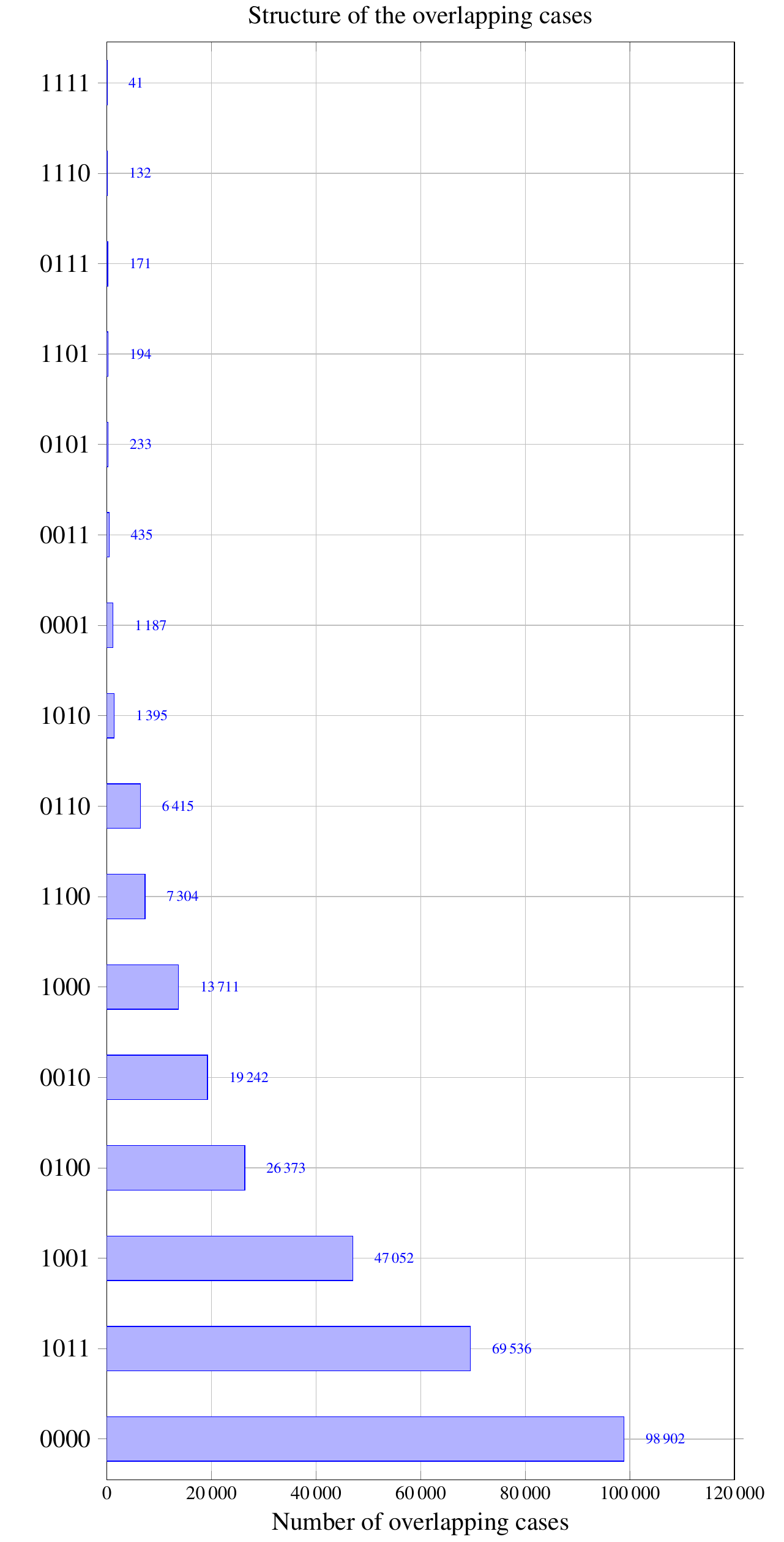}
	\caption{A structure of four-digit classification for Lower Saxony healthcare facilities data from years 2008-2015.\label{fig:overlap:comp}}
\end{figure}

The results of our classification is presented in Table~\ref{tab:overlap:code} and in Figure~\ref{fig:overlap:comp}. 
The most frequently appearing overlap groups for the Lower Saxony healthcare facilities characterized by different declared hospitals are 0000 (different: facility, diagnose, admission and discharge date) 98\,902 detected cases and 0100 (similar as 0000, but with the same diagnose) --  26\,373 cases. Overlaps with the same admission dates but with different discharge dates are also quite frequent 0010 and 0110 -- 19\,242 and 6\,415 cases, respectively.

In case of overlaps taking place in the same facility (first digit in 0-1 classification equal to 1), the most numerous cases are 1011 (different diagnosis the same admission and discharge date), 1001 (different diagnosis and admission dates, the same discharge date) and  1000 (different: diagnosis, admission and discharge dates) resulting in 69\,536, 47\,053 and 13\,711 detected cases, respectively.


\begin{table}
	\centering
	\caption{Effect of four-digit categorisation of the overlapping cases for Lower Saxony healthcare facilities.\label{tab:overlap:code}}
	\input{_khbula03jednoczesne_podwojne_ilosci_kody_sort_a.tex}
\end{table}

Each of four-digit set has been analysed further by assigning each diagnose into groups indexed by numbers according to the rules presented in Table~\ref{tab:diagno}.
%
In Table~\ref{tab:overlap:cody} we summarise the most frequently appearing diagnosis within the particular types of overlaps.
From the presented results we see that most frequent diagnosis for two overlapping records related to different healthcare facilities is disease of the circulatory system [9, 9]. Mental disorders [5, 5] or injuries [19, 19] are also the frequent problem in this case. 

%
For the overlaps characterized by the same healthcare facility records the vast majority of cases are [15, $x$], related to birth and pregnancy. Indeed, the majority of overlapping records coming form the same facility is related to the fact that infants do not posses its own patient identification number and they are registered in the system under the identification number of one of the parents. Second, but much less abundant group of the same healthcare facility overlaps are [5, $x$] --- mental disorders.

\begin{landscape}
	\begin{table}
		\caption{Description of disease indices (ICD 10 code list: \url{http://www.icd10data.com/ICD10CM/Codes}) \label{tab:diagno}}
		\begin{tabular}{ r | r | l}
			Index & ICD 10 & Description \\
			\hline
			1 & A,B &  Infectious and parasitic diseases \\
			2 & C,D &  Neoplasms \\
			3 & D   &  Diseases of the blood and blood-forming organs and certain disorders involving the immune mechanism\\
			4 & E   &  Endocrine, nutritional and metabolic diseases\\
			5 & F   &  Mental, Behavioral and Neurodevelopmental disorders \\
			6 & G   &  Diseases of the nervous system \\
			7 & H   &  Diseases of the eye and adnexa\\
			8 & H   &  Diseases of the ear and mastoid process \\
			9 & I   &  Diseases of the circulatory system \\
			10 & J   &  Diseases of the respiratory system\\
			11 & K   &  Diseases of the digestive system\\
			12 & L   &  Diseases of the skin and subcutaneous tissue\\
			13 & M   &  Diseases of the musculoskeletal system and connective tissue\\
			14 & N   &  Diseases of the genitourinary system \\
			15 & O   &  Pregnancy, childbirth and the puerperium\\
			16 & P   &  Certain conditions originating in the perinatal period\\
			17 & Q   &  Congenital malformations, deformations and chromosomal abnormalities\\
			18 & R   &  Symptoms, signs and abnormal clinical and laboratory findings, not elsewhere classified\\
			19 & S,T &  Injury, poisoning and certain other consequences of external causes\\
			21 & Z   &  Factors influencing health status and contact with health services\\
		\end{tabular}
	\end{table}
\end{landscape}

\begin{landscape}
\begin{table}
	\caption{Number of cases for a given diagnosis (described by two numbers, see Table~\ref{tab:diagno} for reference) for particular groups of overlaps. Two record overlaps are included in this table (vast majority among all overlaps). In square brackets we provide diagnose codes for overlaps versus number of cases. \label{tab:overlap:cody}}                                              
	\scalebox{0.92}{ 
		\input{_khbula03jednoczesne_podwojne_ilosci_kody_sort_b.tex}
		}
\end{table}
\end{landscape}

\section{Towards hospital transfer network}
\label{sec:network}

	Aim of analysis of the overlapping records is to provide model, allowing simulation of patient movement between healthcare facilities and the disease transmission in German healthcare system. We distinguish between two kinds of patient transfers: direct transfer between healthcare facilities (without stay at home) and indirect transfer (otherwise). The question is, whether it is possible to reliably deduce both types of transfers with the data we already have, or is more information necessary.
	
	The indirect transfers are generally easy to determine in this setting, as we can simply trace stay-at-home periods between hospitalisations. The only problematic cases here are when a patient is dismissed from or admitted to many facilities at the same day, but these situations do not occur frequently. Among the problematic overlaps for Lower Saxony healthcare facilities presented in~Table~\ref{tab:overlaps}, the problematic cases correspond to first/last day transfers and admissions to many institutions (overlaps type: 4 and 6-10). These cases constitute about 10\% of total number of overlaps, and taking into account the fact that there are more than 4 million records in total, they are too scarce to considerably impact the results.

	In case of direct transfer, the situation is as follows. Admissions to single institutions (more than 46\% of overlaps) do not contribute to direct transfers. \emph{Standard transfers} and \emph{temporary transfers} are the obvious cases. Here also \emph{first/last day transfers} can be resolved easily, as we can simply assume that there is a single transfer from/to the one-day hospitalisation facility at beginning/end of the hospitalisation.
	Therefore we are left with less than 2\% of problematic cases, where some of them may still be resolved (cf. Figure~\ref{fig:examp}).

	Thus, we conclude that the data provided by AOK Lower Saxon can be used to deduce the direct/indirect transfer rate between hospitals in Lower Saxony. The next step will be to build the transfer network with the resulting transfers. 

\section{Acknowledgements}
This work was supported by grant no.~2016/22/Z/ST1/00690 of National Science Centre, Poland within the transnational research programme JPI-EC-AMR (Joint Programming Initiative on Antimicrobial Resistance) entitled "Effectiveness of infection control strategies against intra- and inter-hospital transmission of MultidruG-resistant Enterobacteriaceae – insights from a multi-level mathematical NeTwork model" (EMerGe-Net). 

We thank the AOK Lower Saxony for providing anonymized record data.

%
%

\bibliographystyle{siam}
\bibliography{literature}

\end{document}

%% file: _khbula03jednoczesne_podwojne_ilosci_kody_sort_a.tex
\begin{tabular}{|l|c|c|c|c|c|c|c|c|}
	\hline
	{ {\bf overlap code}} & { 0000} & { 1011} & { 1001} & { 0100} & { 0010} & { 1000} & { 1100} & { 0110}\\
	\hline
	{ {\bf \# cases}} & { 98902} & { 69536} & { 47052} & { 26373} & { 19242} & { 13711} & { 7304} & { 6415}\\
	\hline
	\hline
	{ {\bf overlap code}} & { 1010} & { 0001} & { 0011} & { 0101} & { 1101} & { 0111} & { 1110} & { 1111}\\
	\hline
	{ {\bf \# cases}} & { 1395} & { 1187} & { 435} & { 233} & { 194} & { 171} & { 132} & { 41}\\
	\hline
\end{tabular}

%% file: _khbula03jednoczesne_podwojne_ilosci_kody_sort_b.tex
\begin{tabular}{|l|r | |l|r | |l|r | |l|r | |l|r | |l|r | |l|r | |l|r | |}
	\hline
	{ 0000} & { Over.} & { 0001} & { Over.} & { 0010} & { Over.} & { 0011} & { Over.} & { 0100} & { Over.} & { 0101} & { Over.} & { 0110} & { Over.} & { 0111} & { Over.}\\
	\hline
	{ [9, 9]} & { 19509} & { [5, 5]} & { 194} & { [9, 9]} & { 4412} & { [5, 5]} & { 80} & { [9, 9]} & { 12107} & { [9, 9]} & { 97} & { [9, 9]} & { 3223} & { [9, 9]} & { 62}\\
	\hline
	{ [5, 5]} & { 6568} & { [9, 9]} & { 155} & { [5, 5]} & { 2319} & { [9, 9]} & { 79} & { [19, 19]} & { 3355} & { [5, 5]} & { 50} & { [19, 19]} & { 907} & { [5, 5]} & { 48}\\
	\hline
	{ [5, 19]} & { 5210} & { [5, 19]} & { 100} & { [19, 19]} & { 1090} & { [5, 19]} & { 39} & { [2, 2]} & { 2642} & { [2, 2]} & { 19} & { [5, 5]} & { 453} & { [19, 19]} & { 15}\\
	\hline
	{ [2, 2]} & { 4170} & { [5, 18]} & { 53} & { [14, 14]} & { 939} & { [6, 18]} & { 19} & { [5, 5]} & { 1860} & { [19, 19]} & { 14} & { [14, 14]} & { 387} & { [18, 18]} & { 12}\\
	\hline
	{ [19, 19]} & { 2946} & { [9, 18]} & { 43} & { [5, 19]} & { 782} & { [5, 18]} & { 18} & { [11, 11]} & { 1073} & { [10, 10]} & { 10} & { [11, 11]} & { 321} & { [6, 6]} & { 11}\\
	\hline
	{ [5, 9]} & { 2608} & { [5, 9]} & { 40} & { [15, 15]} & { 725} & { [9, 18]} & { 17} & { [14, 14]} & { 958} & { [14, 14]} & { 8} & { [15, 15]} & { 217} & { [15, 15]} & { 5}\\
	\hline
	{ [10, 10]} & { 2525} & { [5, 11]} & { 29} & { [11, 11]} & { 695} & { [18, 18]} & { 15} & { [13, 13]} & { 953} & { [13, 13]} & { 7} & { [18, 18]} & { 178} & { [14, 14]} & { 5}\\
	\hline
	{ [14, 14]} & { 2282} & { [2, 2]} & { 29} & { [6, 9]} & { 539} & { [14, 14]} & { 14} & { [10, 10]} & { 857} & { [11, 11]} & { 6} & { [10, 10]} & { 144} & { [11, 11]} & { 4}\\
	\hline
	{ [5, 18]} & { 2189} & { [5, 6]} & { 26} & { [6, 6]} & { 434} & { [6, 6]} & { 13} & { [6, 6]} & { 601} & { [18, 18]} & { 5} & { [6, 6]} & { 132} & { [17, 17]} & { 2}\\
	\hline
	{ [11, 11]} & { 2179} & { [14, 14]} & { 24} & { [10, 10]} & { 434} & { [5, 6]} & { 13} & { [1, 1]} & { 363} & { [6, 6]} & { 5} & { [2, 2]} & { 103} & { [10, 10]} & { 2}\\
	\hline
	\hline
	{ 1000} & { Over.} & { 1001} & { Over.} & { 1010} & { Over.} & { 1011} & { Over.} & { 1100} & { Over.} & { 1101} & { Over.} & { 1110} & { Over.} & { 1111} & { Over.}\\
	\hline
	{ [5, 5]} & { 3877} & { [15, 21]} & { 41482} & { [15, 21]} & { 288} & { [15, 21]} & { 63052} & { [5, 5]} & { 5539} & { [5, 5]} & { 81} & { [5, 5]} & { 71} & { [16, 16]} & { 23}\\
	\hline
	{ [5, 19]} & { 1200} & { [15, 16]} & { 4814} & { [15, 16]} & { 271} & { [15, 16]} & { 5760} & { [2, 2]} & { 1034} & { [10, 10]} & { 41} & { [2, 2]} & { 22} & { [21, 21]} & { 5}\\
	\hline
	{ [5, 9]} & { 653} & { [15, 17]} & { 344} & { [5, 5]} & { 192} & { [15, 17]} & { 516} & { [6, 6]} & { 132} & { [2, 2]} & { 35} & { [15, 15]} & { 17} & { [15, 15]} & { 4}\\
	\hline
	{ [5, 11]} & { 525} & { [5, 5]} & { 52} & { [5, 19]} & { 161} & { [15, 18]} & { 44} & { [9, 9]} & { 118} & { [13, 13]} & { 8} & { [9, 9]} & { 6} & { [5, 5]} & { 3}\\
	\hline
	{ [2, 2]} & { 494} & { [5, 19]} & { 41} & { [15, 15]} & { 80} & { [21, 21]} & { 40} & { [15, 15]} & { 89} & { [9, 9]} & { 6} & { [12, 12]} & { 5} & { [11, 11]} & { 1}\\
	\hline
	{ [5, 18]} & { 468} & { [15, 18]} & { 33} & { [14, 21]} & { 55} & { [8, 15]} & { 24} & { [13, 13]} & { 80} & { [8, 8]} & { 5} & { [16, 16]} & { 2} & { [12, 12]} & { 1}\\
	\hline
	{ [5, 6]} & { 396} & { [10, 21]} & { 28} & { [5, 18]} & { 46} & { [2, 15]} & { 13} & { [19, 19]} & { 50} & { [15, 15]} & { 5} & { [6, 6]} & { 2} & { [14, 14]} & { 1}\\
	\hline
	{ [5, 10]} & { 359} & { [5, 18]} & { 18} & { [5, 6]} & { 37} & { [16, 21]} & { 8} & { [11, 11]} & { 45} & { [4, 4]} & { 3} & { [14, 14]} & { 2} & { [2, 2]} & { 1}\\
	\hline
	{ [4, 5]} & { 322} & { [5, 9]} & { 17} & { [5, 9]} & { 37} & { [11, 15]} & { 7} & { [12, 12]} & { 41} & { [18, 18]} & { 3} & { [19, 19]} & { 2} & { [18, 18]} & { 1}\\
	\hline
	{ [9, 9]} & { 232} & { [12, 15]} & { 13} & { [5, 11]} & { 31} & { [12, 15]} & { 7} & { [14, 14]} & { 41} & { [11, 11]} & { 2} & { [3, 3]} & { 1} & { [6, 6]} & { 1}\\
	\hline
\end{tabular}